\documentclass[aps,prd,superscriptaddress,twocolumn,nofootinbib,10pt]{revtex4}
\usepackage[pdftex]{color}
\usepackage[sort&compress]{natbib}
\usepackage[colorlinks=true,linkcolor=blue,filecolor=blue,urlcolor=blue,citecolor=blue,pdftex,plainpages=false]{hyperref}
\usepackage{url}
\pdfoutput=1
\usepackage{ifpdf}
\usepackage{color,hhline,psfrag,rotating}
\usepackage{dcolumn}
\usepackage{verbatim}
\usepackage{bm}
\usepackage{graphicx}
\usepackage{amsfonts,amssymb,amsmath}
\usepackage{booktabs}  

\newcommand{\MSbar}{\overline{\text{MS}}}
\newcommand{\MSbarSMOM}{\overline{\text{MS}}/\text{SMOM}}
\newcommand{\GeV}{\text{ GeV}}
\newcommand{\tuned}{\text{tuned}}
\newcommand{\sea}{\text{sea}}

\newcommand{\Xsl}[1]{\raise.15ex\hbox{/}\kern-.57em #1}
\DeclareMathOperator{\Tr}{Tr} 

\begin{document}
\title{Determination of quark masses from $\mathbf{n_f=4}$ lattice QCD and the RI-SMOM intermediate scheme}

\author{A.~T.~Lytle}
\email[]{andrew.lytle@glasgow.ac.uk}
\affiliation{SUPA, School of Physics and Astronomy, University of Glasgow, Glasgow, G12 8QQ, UK}
\author{C.~T.~H.~Davies}
\email[]{christine.davies@glasgow.ac.uk}
\affiliation{SUPA, School of Physics and Astronomy, University of Glasgow, Glasgow, G12 8QQ, UK}
\author{D.~Hatton}
\affiliation{SUPA, School of Physics and Astronomy, University of Glasgow, Glasgow, G12 8QQ, UK}
\author{G.~P.~Lepage}
\affiliation{Laboratory for Elementary-Particle Physics, Cornell University, Ithaca, New York 14853, USA}
\author{C.~Sturm}
\affiliation{Institut f{\"u}r Theoretische Physik und Astrophysik, Universit{\"a}t W{\"u}rzburg, Emil-Hilb-Weg 22, D-97074 W{\"u}rzburg, Germany}
\collaboration{HPQCD collaboration}
\homepage{http://www.physics.gla.ac.uk/HPQCD}
\noaffiliation

\date{\today}

\begin{abstract}
We determine the charm and strange quark masses 
in the $\MSbar$ scheme,
using $n_f=2+1+1$ lattice QCD calculations with highly
improved staggered quarks (HISQ)
and the RI-SMOM intermediate scheme to connect the 
bare lattice quark masses
to continuum renormalisation schemes.
Our study covers analysis of systematic uncertainties 
from this method, including nonperturbative artefacts and the impact of  
the non-zero physical sea quark 
masses. 
We find $m_c^{\MSbar}(3 \GeV) = 0.9896(61)$ GeV and $m_s^{\MSbar}(3 \GeV) = 0.08536(85)$ GeV, 
where the uncertainties are dominated by the tuning of the bare lattice 
quark masses.
These results are consistent with, and of similar accuracy to, those using 
the current-current correlator approach coupled to high-order continuum 
QCD perturbation theory, implemented in the same quark formalism and 
on the same gauge field configurations. This provides a strong test of the 
consistency of methods for determining the quark masses to high precision 
from lattice QCD. We also give updated lattice QCD world averages for $c$ and $s$ 
quark masses. 
\end{abstract}

\maketitle
\section{Introduction}
\label{sec:intro}
Quark masses are fundamental parameters of the Standard Model
which must be connected via theory to experimentally measured quantities.
They arise in the Standard Model from interactions with the Higgs
field, and precise knowledge of quark masses will be needed to
test stringently
the Standard Model picture of mass generation~\cite{Lepage:2014fla}. 

In lattice QCD simulations the
bare quark masses of the theory are input parameters, 
and these are tuned to reproduce a
set number of physical observables, typically meson masses (one for each quark mass in the simulation).
These parameters are however defined at the cutoff scale
of the theory and are non-universal, because they depend on the specific lattice 
regularisation of QCD used. To be useful, these values must then be converted 
to a chosen quark mass definition in a continuum regularisation of QCD 
at a fixed physical scale. The conversion, or mass renormalisation, factor 
adjusts for the different treatment of ultraviolet modes on the lattice and 
in the continuum and so in principle can be calculated straightforwardly by a 
`matching' calculation in  
lattice QCD and continuum QCD perturbation theory. 
Lattice QCD perturbation theory~\cite{Mason:2005de} is very hard beyond the first order in the 
strong coupling constant, $\alpha_s$, and 
so this method is limited to an accuracy of several percent~\cite{Mason:2005bj}.  
Higher accuracy can be achieved by methods that make use of nonperturbative 
calculations in lattice QCD combined with continuum QCD perturbation theory 
and we will compare results from two such methods here. 
One issue with these methods is the control of infrared nonperturbative artefacts from the 
lattice QCD calculation that are a source of systematic uncertainty. 

The conventional continuum scheme to which lattice masses are converted is 
the ${\overline{\text{MS}}}$
scheme and we will denote masses in the ${\overline{\text{MS}}}$ scheme
by $\overline{m}$. A scale for the mass must also be chosen and we will use 3 GeV.    
Having a fixed convention for quoting quark masses 
allows a comparison between different determinations. 

One way to make the lattice QCD to continuum QCD quark mass connection is to calculate 
short distance physical quantities in lattice QCD that are 
both sensitive to the quark mass and for which continuum QCD perturbation 
theory (in the $\overline{\text{MS}}$ scheme) has been done to a high order. 
The appropriate energy scale for $\alpha_s$ should also be large.  
A successful method of this type is the `current-current correlator method'~\cite{Allison:2008xk} 
that uses time-moments of heavyonium correlators, extrapolated to the continuum 
from lattice QCD and then matched to QCD perturbation theory accurate through 
$\mathcal{O}(\alpha_s^3)$~\cite{qcdpt1, qcdpt2, Sturm:2008eb, qcdpt3, qcdpt4, qcdpt5}.  
The advantage of this method (which we will denote the JJ method) 
is that nonperturbative effects (condensate contributions), that would 
otherwise obscure the match to perturbation theory, are 
suppressed by four powers of $\Lambda/(2m_h)$, where $m_h$ 
is the heavy quark mass~\cite{Kuhn:2007vp, Chakraborty:2014aca}. 
The suppression is very effective, to the point where these effects have negligible 
impact, because: $\Lambda$ is small 
at around 0.3 GeV; $m_h$ is large (and can be varied to test
the contribution) and 4 is a high power. 
Here $(\Lambda)^4$ represents the expected size of the gluon 
condensate $\langle 0 | \alpha_s G^2/\pi | 0 \rangle$ 
constructed from the gluon field-strength tensor. 

Uncertainties in the JJ method arise from missing higher orders in QCD perturbation theory, but 
these can be tested by implementing the perturbation theory at different scales~\cite{Chakraborty:2014aca}. 
This method has given 1\% accurate results for charm and bottom quark 
masses in 
the $\overline{\text{MS}}$ 
scheme~\cite{Allison:2008xk, McNeile:2010ji, Colquhoun:2014ica, Chakraborty:2014aca, Nakayama:2016atf, Maezawa:2016vgv}. 
The results for $\overline{m}_c$ and $\overline{m}_b$ can then be leveraged into an accurate 
result for lighter quark masses, such as the strange quark mass $\overline{m}_s$. 
This is done by determining fully nonperturbatively 
in lattice QCD the ratio of two quark masses, such as $m_c/m_s$, using the same quark 
formalism for both quarks~\cite{Davies:2009ih, Blossier:2010cr, Durr:2011ed, Carrasco:2014cwa, Bazavov:2014wgs, Chakraborty:2014aca}. 
This ratio (in the continuum limit) is independent of the lattice quark formalism or 
continuum scheme and so also holds for the $\overline{\text{MS}}$ scheme at a fixed scale $\mu$. 
Combining the value of the ratio $m_c/m_s$ (which can now be obtained to an accuracy of 
better than 1\%~\cite{Bazavov:2014wgs, Chakraborty:2014aca, Bazavov:2018omf}) with the value for 
$\overline{m}_c$ then yields a 1\% accurate result for $\overline{m}_s$. 
Further ratios between strange and up/down 
quark masses (see, for example,~\cite{Bazavov:2014wgs}) 
can be used to cascade this accuracy down to even lighter quarks. 

Since the JJ method enables the value of the quark mass in the 
$\MSbar$ scheme to be obtained for an input tuned lattice quark mass, 
it is equivalent to (indirectly) determining the mass renormalisation factor, $Z_m^{\MSbar}(\mu)$, 
that connects the two masses~\cite{Chakraborty:2014aca}. 

Another completely different method for making the connection between lattice and 
$\overline{\text{MS}}$ masses is to determine ratios of appropriate matrix elements 
between external quark states of large virtuality, $\mu^2$, that can be calculated 
both in lattice QCD and in the $\overline{\text{MS}}$ scheme in continuum QCD 
perturbation theory~\cite{Martinelli:1994ty}.  
Such calculations must be done in a fixed gauge, usually Landau gauge. 
The method proceeds by imposing `momentum-subtraction' renormalisation 
conditions~\cite{Sturm:2009kb} on matrix elements in the lattice QCD calculation. 
e.g. 
\begin{equation}
\label{eq:nprdef}
\left. Z_{\Gamma}\langle p_1 | O_{\Gamma} | p_2 \rangle \right|_{p_1^2=p_2^2=q^2=-\mu^2} = \langle p_1 | O_{\Gamma} | p_2 \rangle_0
\end{equation}
defines $Z_{\Gamma}$ for operator $O_{\Gamma} = \overline{\psi} \Gamma \psi$, where 
$\langle p | O_{\Gamma} | p \rangle_0$ is the tree-level matrix element and $\langle p_1 |$ and 
$| p_2 \rangle$ are external quark states. 
The symmetric kinematic configuration specified here (with $q=p_1-p_2$) corresponds to the 
symmetric momentum-subtraction or SMOM scheme. The importance of 
this configuration will be discussed further below.
Applying the condition of eq.~(\ref{eq:nprdef}) to a scalar operator 
(along with a determination of the wavefunction renormalisation factor) 
gives directly a mass 
renormalisation factor, $Z_m^{\text{SMOM}}(\mu)$, that converts the 
lattice quark mass to that in the SMOM scheme.
Because the SMOM scheme can be implemented in the continuum it can 
itself then be matched to the $\overline{\text{MS}}$ scheme using 
continuum QCD perturbation theory (in the same 
gauge)~\cite{Gorbahn:2010bf, Almeida:2010ns}. Multiplying the lattice bare quark 
mass by the final $Z_m^{\overline{\text{MS}}}(\mu)=Z_m^{\overline{\text{MS}}/\text{SMOM}}(\mu)\times Z_m^{\text{SMOM}}(\mu)$ gives the required $\overline{m}(\mu)$. 
This method has been widely applied to operator renormalisation 
in general and not just the determination of $Z_m$, going under the name 
of the `RI-SMOM' (regularisation-independent 
symmetric momentum-subtraction) scheme~\cite{Sturm:2009kb}. 
For a review of this and the earlier RI-MOM scheme, see~\cite{Aoki:2010yq}. 

The RI-SMOM scheme is expected to work in a window in which 
\begin{equation}
\label{eq:nprwindow}
\Lambda_{\text{QCD}} \ll \mu \ll \frac{\pi}{a} .
\end{equation}
Here the upper limit $a\mu \ll 1$ keeps control of lattice discretisation 
effects and the lower limit guards against being dominated by potentially large 
nonperturbative effects~\cite{Politzer:1976tv} that behave as condensates 
multiplied by inverse powers 
of $\mu$. 
Nonperturbative effects were a major issue with the original 
RI-MOM scheme~\cite{Martinelli:1994ty} which set up the kinematics 
for eq.~(\ref{eq:nprdef}) 
so that $p_1^2=p_2^2=-\mu^2$, but 
$p_1=p_2$ so that $q^2=0$. This `exceptional' configuration gave
rise to
differences, inversely proportional 
to $\mu^2$, between renormalisation 
factors 
that should be the same from chiral symmetry (such as those of the pseudoscalar 
and scalar operators). 
This was coupled in some cases to strong nonperturbative dependence 
of the renormalisation factors on the quark mass, 
see 
for example~\cite{Crisafulli:1997ic, Gimenez:1998ue, Blum:2001sr, Aoki:2007xm}. 

In contrast, since none of the momenta are light-like in the RI-SMOM 
scheme, the operators associated with it can be analysed within the Operator 
Product Expansion (OPE) and sensitivity to nonperturbative effects is under 
better control.
Those associated with spontaneous chiral symmetry breaking, for example, are
more benign, with behaviour as $1/\mu^6$ following expectations 
from the OPE~\cite{Aoki:2007xm, Aoki:2010dy}.  
The SMOM vertex functions show only small quark mass dependence. 
An added bonus is that the RI-SMOM to $\MSbar$ matching 
factors~\cite{Gorbahn:2010bf, Almeida:2010ns} for $Z_m$ are much closer
to unity (through $\mathcal{O}(\alpha_s^2)$) than their 
RI-MOM counterparts~\cite{Franco:1998bm, Chetyrkin:1999pq}.
This means that the RI-SMOM mass renormalisation factor can 
be obtained with smaller systematic uncertainty. 

Nonperturbative 
condensate effects are still present in the RI-SMOM scheme, 
however, and their effects must be included in any accurate 
determination of the quark mass. The leading 
condensate contribution to $Z_m$ is chirally-symmetric 
and is only suppressed 
by $1/\mu^2$. Since the associated condensate is the Landau gauge gluon 
condensate (also known as the gluon mass condesate)~\cite{Chetyrkin:2009kh}, 
$\langle 0 | A^2 | 0 \rangle$, which is thought 
to be $\mathcal{O}(1) \mathrm{GeV}^2$~\cite{Blossier:2010ky, Burger:2012ti}, 
this contribution could have a significant effect up to very high 
values of $\mu^2$. 
Such a contribution must be included in the analysis and constrained with results 
at multiple $\mu$ values. Here we provide 
a thorough analysis of systematic uncertainties in 
the determination of the quark mass with this method, including that of 
nonperturbative effects. 

Using the RI-SMOM intermediate scheme we are then able to 
determine values for $\overline{m}_c$ and $\overline{m}_s$ with 
comparable accuracy, around 1\%, to that obtained using the 
current-current correlator method, and using the same lattice quark 
formalism (highly-improved staggered quarks (HISQ)). The RI-SMOM approach has 
completely different systematic uncertainties, however,  
so that a comparison of results from the 
two methods is then a strong test of our understanding of 
systematic uncertainties, because the lattice bare 
quark masses are tuned to the same values in both cases. 

The paper is laid out as follows: Section~\ref{sec:rismom} 
describes briefly the RI-SMOM approach and Section~\ref{sec:smomstagg} 
gives some details needed to 
implement it for staggered quarks; Section~\ref{sec:lattice} 
then gives results for the lattice determination of $Z_m$ in the 
SMOM scheme; Section~\ref{sec:mass} uses these results to determine 
the quark masses in the $\overline{\text{MS}}$ scheme. 
Finally Section~\ref{sec:conclusions} compares to earlier values, 
giving new world averages, and concludes with prospects for future 
improvements. 

\section{The RI-SMOM method} 
\label{sec:rismom}

As outlined in Section~\ref{sec:intro} the lattice QCD RI-SMOM approach 
mimics what would be done in continuum QCD in a momentum-subtraction scheme. 
A key part of the argument is that the calculation should be set 
up in a way that is regularisation-independent. Thus within the lattice 
QCD calculation the same answer for the quark mass in the SMOM scheme should 
be obtained in any quark formalism up to discretisation effects. 
Then the continuum limit of the lattice result also holds 
in the equivalent continuum SMOM scheme. 
The continuum SMOM to $\MSbar$ matching completes the 
conversion to the $\MSbar$ scheme. Within the lattice QCD 
calculation we must then also ensure that the tuning of quark masses 
and the determination of the lattice spacing are done in a regularisation-independent 
way. This is of course the standard practice when we determine the 
lattice spacing and tune lattice quark masses using physical quantities 
(such as hadron masses) calculated at the lattice QCD physical point 
(i.e. including sea quarks with physical masses) and take the value from experiment.  
We will return to this point below.
 
To determine the renormalisation factor for an operator $O_{\Gamma}$ 
in this framework 
we then need to apply renormalisation conditions to the inverse propagator 
(to obtain a wavefunction renormalisation factor) and to an amputated 
vertex function containing $O_{\Gamma}$.  

For free quarks in the continuum the inverse of the quark propagator, $S(p)$, is 
\begin{equation}
S_0^{-1}(p) = m - \Xsl{p}
\end{equation}
The wavefunction renormalisation factor, $Z_q$, in this 
scheme can be defined by
~\cite{Martinelli:1994ty, Sturm:2009kb} 
\begin{equation}
\label{eq:Zq}
\frac{1}{12p^2} \Tr[S^{-1}(p) \, \Xsl{p}] = -Z_q \\
\end{equation}
so that $Z_q=1$ in the free theory. 

Vertex functions $G_{\Gamma}$ of operator $O_{\Gamma}$ (=$\overline{\psi}\Gamma\psi$) can be calculated between 
two external, off-shell quark lines and 
`amputated' as:
\begin{equation}
\label{eq:lambda}
\Lambda_{\Gamma} = S^{-1}(p_2)G_{\Gamma}S^{-1}(p_1) \, .
\end{equation}
The renormalisation condition (eq.~(\ref{eq:nprdef})) on $\Lambda_{\Gamma}$ 
yields $Z_{\Gamma}/Z_q$ given lattice values for $\Lambda_{\Gamma}$. 
From this we can determine $Z_{\Gamma}$ if we have $Z_q$. 
Here we are interested in the mass renormalisation 
factor, $Z_m = 1/Z_S$ obtained from the scalar quark bilinear:
\newcommand{\sym}{\text{sym}}
\begin{equation}
\label{eq:ZSdef}
\frac{1}{12} \frac{Z_S}{Z_q}
\Tr[\Lambda_{S}(p_1,p_2)]|_\sym = 1  \, .
\end{equation}
Again the tree-level value of $Z_S$ is 1. 
Here $|_\sym$ indicates that
$p_1$ and $p_2$ satisfy $p_1^2 = p_2^2 = (p_1-p_2)^2 = -\mu^2$ 
(the RI-SMOM condition), so that there is
a single momentum scale. 
We will also be interested in the pseudoscalar 
operator with renormalisation condition 
\begin{equation}
\label{eq:ZPdef}
\frac{1}{12i} \frac{Z_P}{Z_q}
\Tr[\Lambda_{P}(p_1,p_2) \gamma_5]|_\sym = 1 \,. 
\end{equation}

This method is straightforward to implement in lattice 
QCD. The inverse propagators and vertex functions are calculated
from ensemble averages over a set of gluon fields. Note that 
this means that eq.~(\ref{eq:lambda}) gives $\Lambda_{\Gamma}$ as 
the product of three ensemble averages. 
$Z_P$ and $Z_S$ in eqs.~(\ref{eq:ZSdef}) and~(\ref{eq:ZPdef})
are then defined as a ratio of ensemble averages, with uncertainties 
determined via a bootstrap procedure. 
In practice, a relatively small number of gluon 
field configurations are needed for good numerical precision in the renormalisation 
factors $Z_S$ and $Z_P$.  

Calculations can readily be done for a range of different masses for 
the `valence' quarks for which propagators are calculated. 
We will use the same quark mass for the two sides of the vertex function 
but note that only quark-line connected Wick contractions appear in 
this calculation. 
It is conventional to define the RI-SMOM renormalisation constants 
in the limit of zero valence quark mass, and we do that here. 
One reason for doing this is that of consistency, since 
the perturbative calculations that match RI-SMOM to $\overline{\text{MS}}$ 
have been done for massless quarks\footnote{Note that it is perfectly possible to define 
an RI-SMOM scheme for nonzero quark mass and match this perturbatively 
to $\overline{\text{MS}}$~\cite{Boyle:2016wis}}. This is discussed further below.  

In practice a more important issue is that 
of nonperturbative quark mass dependence associated with condensate contributions. 
An operator product expansion (OPE) 
approach to the RI-SMOM scheme (where it can be rigorously 
applied) shows that there are contributions to the quark propagators and 
vertex functions used to define $Z_m$ that appear 
as inverse powers of $\mu^2$ multiplied by powers of quark masses, or 
quark or gluon condensates or combinations of all of these~\cite{Aoki:2007xm, Chetyrkin:2009kh}. 
It is important to remember that, because we are dealing with gauge-noninvariant 
quantities here, gauge-noninvariant condensates can also appear. 
These nonperturbative contributions are not part of the
perturbative mass renormalisation factor, but they cannot be trivially 
separated from it in a lattice QCD calculation. 
Although the nonperturbative terms seen in the RI-SMOM scheme are 
well-behaved,
they are not entirely negligible 
at the values of $\mu^2$ that we use here, as we will discuss in
Section~\ref{sec:lattice}. It therefore makes sense 
to remove them, where possible, by extrapolating in the valence quark mass 
to zero. This only works, of course, for cases where the effect is proportional to 
a power of the quark mass (and we will study these in Section~\ref{subsec:valmass}). 
The leading contribution to $Z_m$ in terms of inverse powers of $\mu$ comes 
from the Landau gauge gluon condensate with
no powers of quark masses multiplying it and so it cannot be removed by extrapolating 
to zero quark mass. There are also higher order contributions of this form. 
This means that we have to allow for contributions of this kind in our fit 
ansatz for $Z_m$ and test for them by 
varying $\mu$. This enables us to remove them from our 
determination of the $\overline{\text{MS}}$ quark 
mass and to allow an appropriate uncertainty in our error budget from our incomplete 
knowledge of these contributions. 

Note that the sea quark masses are not extrapolated to 
zero. We use calculations 
at physical values of the masses of the $u$, $d$, $s$ and $c$ quarks in the 
sea (with $m_u = m_d$) to determine the lattice spacing and tune the valence 
masses~\cite{Chakraborty:2014aca}. We also calculate 
$Z_m$ on multiple gluon configurations with different unphysical values 
of the masses of the sea 
quarks (for a given bare coupling) to test the dependence on these parameters.   
As we show in Section~\ref{subsec:seamass} dependence of $Z_m$ on the sea quark 
masses is much smaller than that on the valence quark mass and barely visible.
Nonperturbative contributions arising from the sea quarks, 
some of which depend on the sea quark masses, 
will be present and we have to estimate a systematic error from that effect.  

We return now to the issue of the perturbative matching to 
$\overline{\text{MS}}$. The renormalisation factor between the RI-SMOM scheme and the $\MSbar$ 
scheme has been worked out through $\mathcal{O}(\alpha_s^2)$ in continuum QCD 
perturbation theory in~\cite{Sturm:2009kb, Gorbahn:2010bf, Almeida:2010ns}. 
Writing this renormalisation factor as 
\begin{equation}
\label{eq:Zmpert}
Z_{m}^{\overline{\text{MS}}/\text{SMOM}}(\mu) = 1+c_{1}\alpha_s^{\overline{\text{MS}}}(\mu) + c_{2} [\alpha_s^{\overline{\text{MS}}}(\mu)]^2+\ldots
\end{equation}
we tabulate the results for $c_1$ and $c_2$ in Table~\ref{tab:mom-msbar}. 
These are calculated at zero (valence and sea) quark mass.

\begin{table}
\caption{ Coefficients $c_1$ multiplying $\alpha_s$~\cite{Sturm:2009kb} and 
$c_2$ multiplying $\alpha_s^2$~\cite{Gorbahn:2010bf, Almeida:2010ns} in the matching from 
the RI-SMOM scheme to the $\overline{\text{MS}}$ 
scheme. Results are given for both $n_f=3$ and $n_f=4$ with all quark 
masses set to zero.   
Note how small these coefficients are. 
The equivalent of $c_1$ for the earlier RI-MOM scheme is -0.424 and 
for $c_2$ with $n_f=3$, -0.769~\cite{Franco:1998bm, Chetyrkin:1999pq}.
}  
\label{tab:mom-msbar}
\begin{ruledtabular}
\begin{tabular}{lllllll}
scheme & $c_1$ & $c_2$ \\
\hline 
RI-SMOM ($n_f$=3) & -0.0514 &  -0.0669     \\
RI-SMOM ($n_f$=4) & -0.0514 &  -0.0415 
\end{tabular}
\end{ruledtabular}
\end{table}

We must also account for systematic errors in the perturbative matching 
in the continuum from our RI-SMOM scheme with non-zero sea quark mass 
to the $\overline{\text{MS}}$ scheme. 
Sea quarks appear first at 
$\mathcal{O}(\alpha_s^2)$ in the matching and the largest effect present in 
our calculation will be for the sea $c$ quark. We estimate the size of 
this effect in Appendix~\ref{appendix:seamass}. This gives an adjustment to $c_2$ 
that we will include when evaluating $Z_m^{\MSbar/\text{SMOM}}$ 
in Section~\ref{sec:mass}.

\section{RI-SMOM with staggered quarks} 
\label{sec:smomstagg}

There are minor complications on the lattice QCD side if a staggered quark formalism 
is used, as here, because of the fermion doubling issue. 
The staggered quark action is derived from a naive transcription 
of the Dirac action onto the lattice in which a rotation is made 
to diagonalise the action in spin-space. 
The spin degree of freedom can then be dropped and the 16 `doublers' 
or tastes of the naive action become 4 tastes in the staggered action. 
To reconstruct the 4-taste 4-spin Dirac field then requires 
combining staggered quark fields, $\chi$, over a $2^4$ 
hypercube~\cite{KlubergStern:1983dg}. 
This has implications for the momentum-space quark field that 
enters into the momentum-subtraction renormalisation 
formalism. The full lattice Brillouin zone, in lattice 
units
\begin{equation}
\label{eq:fullbz}
-\pi \le ap \le \pi
\end{equation}
contains, for staggered quarks, both momentum and 
taste information~\cite{Golterman:1984cy}. 
To separate them we must work in a reduced 
Brillouin zone
\begin{equation}
\label{eq:redbz}
-\pi/2 \le ap^{\prime} \le \pi/2
\end{equation}
with an additional 4-dimensional label for each subzone. Then
\begin{equation}
\label{eq:bzmatch}
ap_{\mu} = ap^{\prime}_{\mu} + \pi B_{\mu}
\end{equation}
with $B_{\mu}$ a 4-dimensional vector of 0s and 1s.
We use the method for staggered quarks 
developed in~\cite{Lytle:2013qoa}, 
and here simply give an overview of that procedure. 

For a given momentum (in lattice units)
$ap^{\prime}$ in the reduced Brillouin zone, 
we invert the staggered Dirac operator on 16 momentum sources of 
the form $e^{ipx}$ with 
$ap = ap^{\prime} + \pi A$, where $A$ is a 4-vector composed of 0s and 1s.
Each of the resulting propagators, $S(y,p)$ where 
$y$ runs over the lattice volume, is Fourier transformed 16 times with
momenta $-ap^{\prime} + \pi B$, with $B$ a 4-vector of the same type as $A$. 
The results are assembled into a propagator 
\begin{equation}
S(ap^{\prime}) \equiv S_{AB}(ap^{\prime}) = S(ap^{\prime}+\pi A, -ap^{\prime}+\pi B) .
\end{equation} 
This is a $48\times 48$ matrix, but we have kept the colour indices 
implicit; the matrix is diagonal in colour space on forming the 
ensemble average over lattice gluon fields. 
The propagator is also a taste-singlet~\cite{Lytle:2013qoa} 
and so has the same properties for the purposes of the SMOM
approach to those for other quark formalisms. 
After averaging over gluon fields the matrix is inverted 
for each 
value of $p^{\prime}$ to obtain the inverse propagator. 

To apply the condition in eq.~(\ref{eq:Zq}) to 
determine $Z_q$ we must multiply by a representation of 
the matrix $\Xsl{p^{\prime}}$ in AB space. Using the notation 
of~\cite{Lytle:2013qoa} this is the matrix 
$\hat{p}^{\prime}_{\mu}\overline{\overline{(\gamma_{\mu}\otimes I)}}$
that is the Fourier transform of the (taste-singlet) derivative term in the 
free inverse propagator. Since this derivative is improved to remove 
$a^2$ discretisation effects for our improved staggered quark action,
we 
take $a\hat{p}^{\prime}_{\mu} = \sin(a{p}^{\prime}_{\mu}) + \sin^3(a{p}^{\prime}_{\mu})/6$ so that $Z_q$ is equal to 1 in the free case. 
$\overline{\overline{(\gamma_{\mu}\otimes I)}}$ is a matrix of 0, 1 and -1 
obtained by tracing over products of gamma matrices as described
in Appendix A of~\cite{Lytle:2013qoa}. 
Then 
\begin{equation}
\label{eq:Zqlatt}
Z_q(p^{\prime}) = -\frac{i}{48}\sum_{\mu}\frac{\hat{p}^{\prime}_{\mu}}{(\hat{p}^{\prime})^2}\Tr\left[ \overline{\overline{(\gamma_{\mu}\otimes I)}}S^{-1}(p^{\prime})\right] .
\end{equation}
The trace is over spin, taste and colour. 

The scalar operator that we use to determine the mass 
renormalisation factor is the local 
taste-singlet operator $\overline{\chi}(x)\chi(x)$.  
The vertex function for this operator is then constructed as
\begin{eqnarray}
\label{eq:GSlatt}
G_{S,AB}(p_1,p_2) &=& \\
&& \hspace{-6.0em}\langle \chi(p^{\prime}_1+\pi A) \left(\sum_x \bar{\chi}(x) \chi(x) e^{i(p^{\prime}_1-p^{\prime}_2)x} \right) 
\bar{\chi}(p^{\prime}_2 + \pi B) \rangle \nonumber \\
&& \hspace{-8.0em} = \frac{1}{n_{\text{cfg}}}\sum_{x,\text{cfg}} S(p^{\prime}_1+\pi A,x)e^{i(p^{\prime}_1 - p^{\prime}_2)x}(-1)^{x} S^{\dag}(p^{\prime}_2+ \pi \tilde{B}, x). \nonumber
\end{eqnarray}
Here $(-1)^x$ is the alternating phase factor over the lattice, $(-1)^{x_1+x_2+x_3+x_4}$. 
$S^{\dag}$ is the hermitian conjugate in colour space and has a permuted $B$ index according to 
$\tilde{B}=B+_2(1,1,1,1)$. 
To apply eq.~(\ref{eq:ZSdef}) we must multiply $G_{S,AB}$ on both sides by the inverse propagator 
to give $\Lambda_{S,AB}$
and again take the trace over spin, taste and colour. For the local taste-singlet scalar this 
gives the simple expression 
\begin{equation}
\label{eq:ZS}
\frac{Z_q}{Z_S} = \frac{1}{48} \Tr \Lambda_S(p^{\prime}) .
\end{equation}
For the local pseudoscalar operator the procedure is identical except that there is no $(-1)^x$ in 
the equivalent of eq.~(\ref{eq:GSlatt}) and in the equivalent of eq.~(\ref{eq:ZS}) multiplication 
by the matrix $\overline{\overline{\gamma_5\otimes \gamma_5}}$ is needed before taking the trace. 
This can be written simply as a $16\times 16$ matrix with a skew-diagonal of 1s. 
$1/Z_S=Z_m$ is then obtained by dividing by $Z_q$. 

\section{Lattice QCD calculation}
\label{sec:lattice}
For this calculation we use ensembles of gluon field configurations 
generated by the MILC
collaboration~\cite{Bazavov:2010ru,Bazavov:2012xda}. 
These include $u$, $d$, $s$ and $c$ quarks in the quark sea, with 
$m_u=m_d=m_l$. The gluon action is fully improved to remove 
discretisation errors through $\mathcal{O}(\alpha_sa^2)$~\cite{Hart:2008sq}. 
The sea quarks are implemented through the Highly Improved Staggered 
Quark (HISQ) formalism~\cite{HISQ, fdshort} designed, and demonstrated, 
to have very small 
discretisation effects, at $\alpha_s^2a^2$ and $a^4$. 
We also use the HISQ formalism for our propagator and vertex function
calculations. 
The simulation parameters for the sets (ensembles) of gluon field 
configurations used are given 
in Table~\ref{tab:ensembles}. We have sets at three different values 
of the bare QCD coupling, $\beta$, with finer lattice spacing as 
$\beta$ increases. For $\beta=6.0$, referred to here as `coarse' lattices,  
we have 7 different values of the sea quark masses, varying over a 
wide range. This enables us to test the dependence on the sea quark 
masses of our results. We also have 3 different values of the lattice 
spatial volume to test for volume-dependence. On `coarse' and `fine' lattices 
we include ensembles with physical sea $u/d$ (as well as $s$ and $c$) quark masses.   

\begin{table}
\caption{Simulation parameters for the MILC gluon field 
ensembles that we use, labelled by set number in the first column. 
$\beta=10/g^2$ is the bare QCD coupling and $L_s$ and $L_t$ 
give the lattice dimensions. $am_l^{\sea}$, $am_s^{\sea}$ 
and $am_c^{\sea}$ give the sea quark masses in lattice units. 
Sets 1--9 will be referred to in the text as `coarse', sets 10 and 11 as `fine' 
and set 12 as `superfine'. Most of the sets we show here are used to 
test dependence on sea quark masses or spatial volume; our final 
analysis will be done using results from sets 2, 4, 9, 10, 11 and 12 
(marked in bold in the table below).  
}  
\label{tab:ensembles}
\begin{ruledtabular}
\begin{tabular}{lllllll}
Set & $\beta$ & $L_s$ & $L_t$ & $am_l^{\sea}$ & $am_s^{\sea}$ & $am_c^{\sea}$ \\
\hline 
1 & 6.0 & 20 & 64 & 0.008 & 0.040 & 0.480 \\
{\bf 2} & 6.0 & 24 & 64 & 0.0102 & 0.0509 & 0.635 \\
3 & 6.0 & 24 & 64 & 0.00507 & 0.0507 & 0.628 \\
{\bf 4} & 6.0 & 32 & 64 & 0.00507 & 0.0507 & 0.628 \\
5 & 6.0 & 40 & 64 & 0.00507 & 0.0507 & 0.628 \\
6 & 6.0 & 32 & 64 & 0.00507 & 0.00507 & 0.628 \\
7 & 6.0 & 32 & 64 & 0.00507 & 0.012675 & 0.628 \\
8 & 6.0 & 32 & 64 & 0.00507 & 0.022815 & 0.628 \\
{\bf 9} & 6.0 & 48 & 64 & 0.00184 & 0.0507 & 0.628 \\
{\bf 10} & 6.30 & 48 & 96 & 0.00363 & 0.0363 & 0.430 \\
{\bf 11} & 6.30 & 64 & 96 & 0.00120 & 0.0363 & 0.432 \\
{\bf 12} & 6.72 & 48 & 144 & 0.0048 & 0.024 & 0.286 \\
\end{tabular}
\end{ruledtabular}
\end{table}

\begin{figure}[ht]
\includegraphics[width=0.47\textwidth]{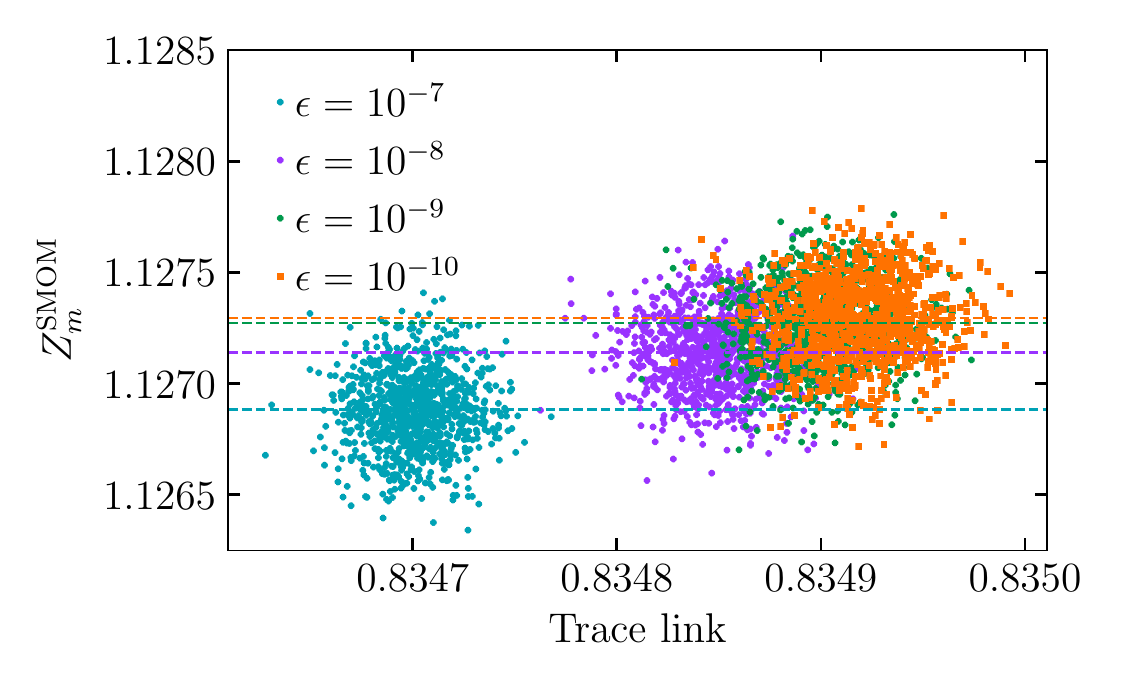}
\caption{ Scatter plots from bootstrap samples of $Z_m^{\mathrm{SMOM}}$ at $\mu$ = 2 GeV 
and Landau gauge Trace link 
values on coarse set 4, for a quark mass value in lattice units of 0.0153. 
On the left results are for a gauge-fixing tolerance on $10^{-7}$ (the tolerance 
we use); on the 
right the tolerance is successively tightened to $10^{-10}$. Mean values for $Z_m^{\mathrm{SMOM}}$ 
are indicated by dashed lines of matching colour. 
}
\label{fig:gfscatter}
\end{figure}

\begin{figure}[ht]
\includegraphics[width=0.47\textwidth]{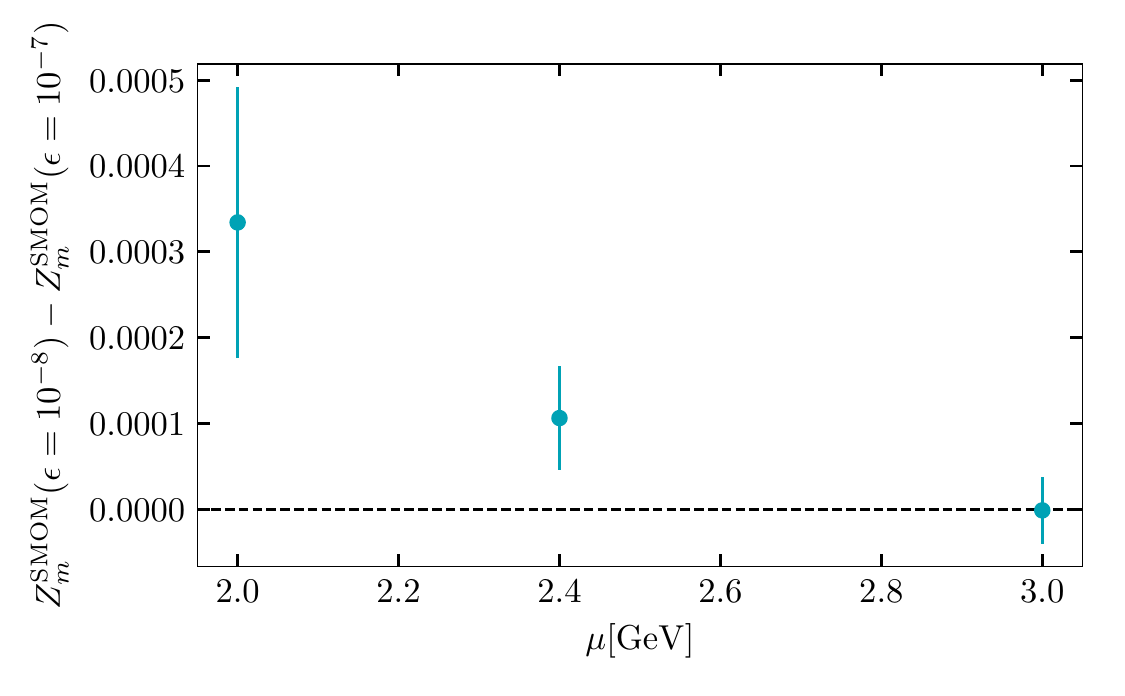}
\caption{ The impact of the gauge fixing tolerance on $Z_m^{\mathrm{SMOM}}(\mu)$ as a function 
of $\mu$. 
Results are for coarse set 2, for a quark mass in lattice units of 0.0204, and show a steep fall 
with value of $\mu$ consistent with sensitivity to gauge-noninvariant condensates. 
}
\label{fig:gfdiffmu}
\end{figure}

We fix the gauge field configurations to lattice Landau gauge by 
maximising the average trace over colour of the gluon field link. 
Note that this differs from the continuum Landau gauge by 
discretisation errors~\cite{Hart:2004jn}. 
On each ensemble we then calculate quark propagators for a range 
of quark masses and momentum values and assemble vertex 
functions as described in Section~\ref{sec:rismom}.
We use the bootstrap method to determine the 
uncertainty in $Z_m^{\mathrm{SMOM}}$ from combining $Z_S$ and $Z_q$, 
as well as the correlations between results obtained on a given 
ensemble.
High precision is possible with these calculations with only a 
moderate number of samples of gluon field configurations. 
We use 20 from each set, well-spaced in Monte Carlo generation time 
for statistical errors below 0.1\%. We have tested that the statistical errors 
are Gaussian by comparing the mean and median from a bootstrap distribution.  
We have also checked that the tolerance we use for 
the Dirac matrix inversion is such that tightening the tolerance has 
no significant effect on the results. 
The impact of the gauge-fixing tolerance will be discussed below. 

For the RI-SMOM calculations reported here, 
the momenta that we use in the two propagators combined in the 
vertex function 
are given in lattice units by
\begin{eqnarray}
\label{eq:pdef}
ap_1^{\prime} &=& \frac{2\pi}{L_s} (x+\frac{\theta}{2}, 0, x+\frac{\theta}{2}, 0) , \\
ap_2^{\prime} &=& \frac{2\pi}{L_s} (x+\frac{\theta}{2}, -x-\frac{\theta}{2}, 0, 0) \nonumber
\end{eqnarray}
for integer $x$.  Adding the additional $\theta/2$ term through 
a `momentum-twist' using phased boundary conditions~\cite{twist, Arthur:2010ht} allows us to 
tune the value of the momenta used precisely. This means, for 
example, that we can tune the momenta to be the same on ensembles 
with different values of $L_s$.   
With the definitions of eq.~(\ref{eq:pdef}) 
$a(p_1^{\prime}-p_2^{\prime})$ has the same magnitude 
as each of $ap_1^{\prime}$ and $ap_2^{\prime}$ which is the 
appropriate kinematics for the RI-SMOM scheme. We will call 
this magnitude $a\mu$: 
\begin{equation}
\label{eq:mudef}
a\mu = a|p_1^{\prime}|=a|p_2^{\prime}|=a|p_1^{\prime}-p_2^{\prime}|. 
\end{equation}
We use momenta 
only in the spatial directions for simplicity because our lattices have a 
different extent in the time direction.
We use a variety of $x$ and $\theta$ values and 
have tested that results do not change under 
a change of $x$ and $\theta$ to achieve the same $ap$. 
Note that, in keeping with eq.~(\ref{eq:redbz}),
we do not want any momentum component in lattice units to exceed $\pi/2$. 
This limits how high a value of $\mu$ we can reach; 
for example we cannot exceed a $\mu$ value of 3 GeV on the coarse lattices.

The results enable us to extract a renormalisation factor for 
the scalar current, $Z_m^{\mathrm{SMOM}}$, in the RI-SMOM scheme for each 
ensemble for a variety of $a\mu$ values and HISQ quark masses used 
in the propagators, $am$. Insofar as $Z_m$ is a renormalisation 
factor from one QCD regularisation scheme to another, taking account 
of the differences in the two schemes at the cutoff, we expect 
$Z_m$ to behave as a power series in $\alpha_s$ with coefficients 
that depend logarithmically on the ratio of the two cutoffs, i.e. on $\ln(a\mu)$.  
Since $Z_m$ here is being determined nonperturbatively in lattice 
QCD, differences from this expectation arise both for small $a\mu$ and 
large $a\mu$ values and we will address both of these here. 

For large $a\mu$ values, systematic discretisation effects can 
appear from the granularity of the lattice. 
In $Z_m$ such effects would cause systematic 
errors of the form $(a\mu)^n$ where $n$ is a positive power 
whose value depends on the quark action, a higher power corresponding 
to a more highly improved action. With the HISQ action 
we have removed tree-level $a^2$ terms and so we expect 
discretisation effects at $(a\mu)^2$ to be suppressed by powers 
of $\alpha_s$ and therefore to be relatively small~\cite{HISQ}. The lowest order at which 
tree-level discretisation errors can appear is at $(a\mu)^4$.   
In fact the discretisation errors, as long as they are not too large,
are benign. In the end, in order to determine a quark 
mass relevant to the physical world, we will perform an 
extrapolation in $a$ to the continuum limit $a = 0$, at fixed $\mu$, and 
remove discretisation errors. 

\begin{figure}[ht]
\includegraphics[width=0.45\textwidth]{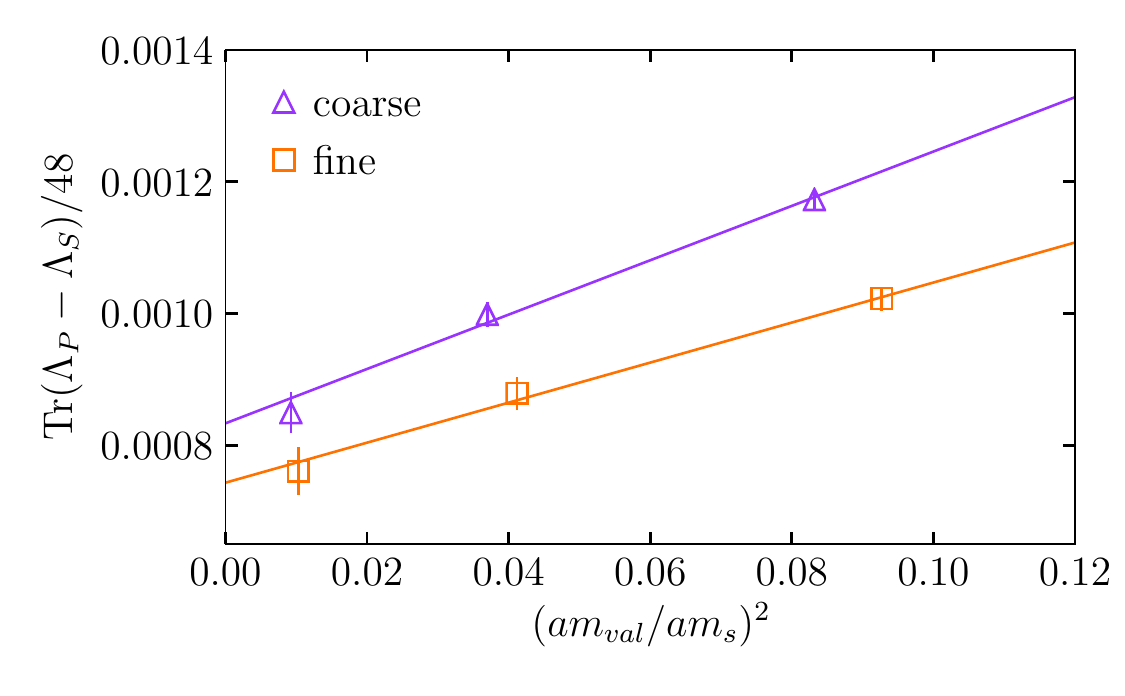}
\includegraphics[width=0.45\textwidth]{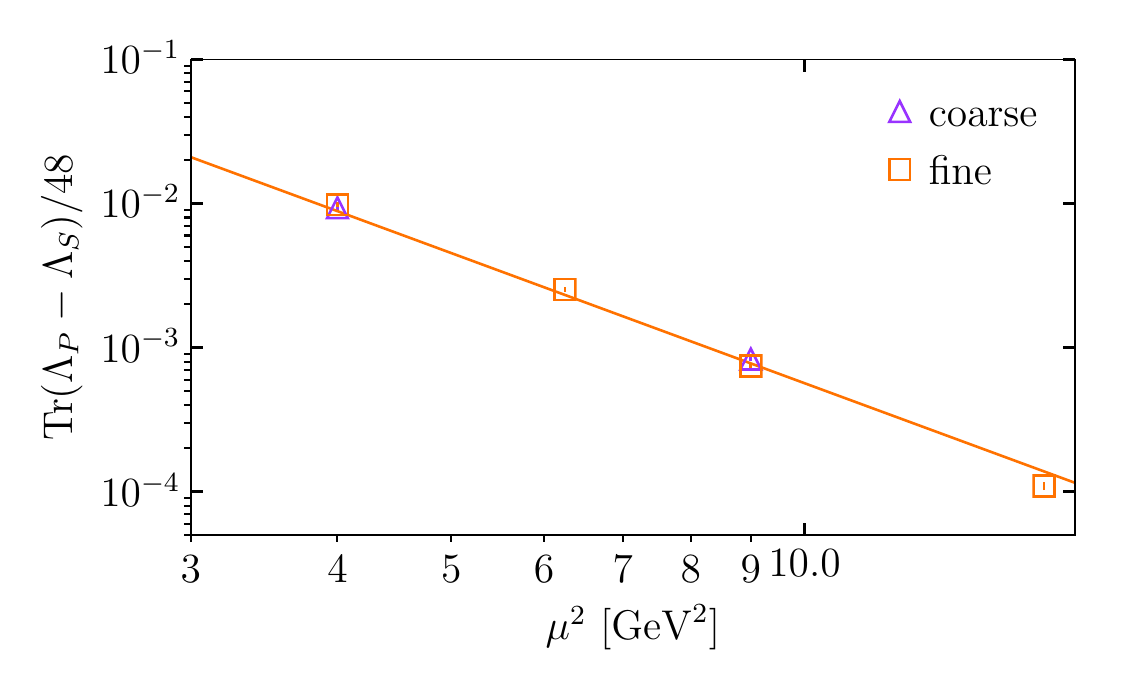}
\caption{ The difference of vertex functions $\Lambda_P$ and 
$\Lambda_S$ (proportional to $Z_P-Z_S$) 
as a function of valence quark mass and as a function of $\mu$. 
Upper plot: $\mathrm{Tr}(\Lambda_P-\Lambda_S)/48$ at $\mu$ = 3 GeV plotted against 
the square of the valence quark mass in units of the tuned $s$ quark mass. 
Results are shown for coarse and fine lattices (sets 3 and 10). 
Lower plot: $\mathrm{Tr}(\Lambda_P-\Lambda_S)/48$ extrapolated to zero valence quark 
mass, following the upper plot, now plotted against $\mu$
on a log-log scale. For comparison, the line shows a constant divided by $\mu^6$. 
This plot shows that, in the RI-SMOM scheme, the nonperturbative contributions 
to $\Lambda_P-\Lambda_s$ are much more strongly 
suppressed than in the RI-MOM scheme, falling as $\mu^{-6}$.  
}
\label{fig:ZPS}
\end{figure}

Of more concern are nonperturbative effects that can have 
an impact
at small values of $\mu$. An operator product expansion (OPE) 
tells us that the vertex functions can be expanded in inverse powers of 
$\mu$ with coefficients that depend on condensates, vacuum expectation 
values of local quark and gluon operators. In the current-current 
correlator method, this effect was studied in~\cite{Chakraborty:2014aca}. 
There the heavy quark mass, $m_h$, replaces $\mu$ and nonperturbative 
terms of the form  
\begin{equation}
\label{eq:cond}
d_1 \frac{m\langle \overline{\psi}\psi \rangle}{(2m_h)^4} + d_2 \frac{\langle \alpha_s G^2/\pi \rangle}{(2m_h)^4} + \ldots
\end{equation}
can appear in the correlator moments. The first term contains the light quark chiral 
condensate $\overline{\psi}\psi$ and the second, the gluon condensate constructed 
from the gluon field-strength tensor (the heavy quark 
condensate being absorbed into this). Since the current-current 
correlator method uses gauge-invariant correlation functions only gauge-invariant 
condensates can appear. To mass dimension four these are the only possibilities. 
The size of such condensates is typically $[\mathcal{O}(300\,\text{MeV})]^n$ where 
$n$ is their mass dimension. 

Here we use gauge-noninvariant vertex functions and propagators and so gauge-noninvariant 
condensates can appear. 
Such condensates can be larger in magnitude than the gauge-invariant ones because 
powers of the Landau gauge gluon field, $A_{\mu}$, can appear (see, for example,~\cite{Blossier:2010ky} 
for the gluon case) and this can be associated with inverse powers of $\mu$ as small 
as $\mu^{-2}$. In Section~\ref{sec:mass} and Appendix~\ref{appendix:ope} we 
discuss how we expect such 
a condensate to affect $Z_m$.    

Study of the impact of the gauge-fixing tolerance provides some evidence of 
sensitivity to these gauge-noninvariant nonperturbative effects.  
For the Landau gauge-fixing, we use a tolerance of $10^{-7}$ 
on the magnitude of the gradient of the gauge field on gluon configuration 
sets 1 through 11. This 
fixes the average trace of the link in Landau gauge to a few parts in 
10,000. The residual effect on $Z_m$ from this gauge-fixing tolerance 
is at the same level as we now demonstrate.
Figure~\ref{fig:gfscatter} shows a scatter plot of bootstrap samples for
$Z_m^{\mathrm{SMOM}}$ on coarse set 4 at $\mu$ = 2 GeV for a gauge-fixing tolerance of $10^{-7}$ 
and then successive tightening of this tolerance by factors of 10 down 
to $10^{-10}$. The tighter tolerance gives a shift in the mean value of $Z_m^{\mathrm{SMOM}}$ by 
around 0.0004. Results at higher $\mu$ values show much  
smaller effects in a way that demonstrates their origin in nonperturbative effects. 
This is illustrated in Figure~\ref{fig:gfdiffmu} which shows the change in $Z_m^{\mathrm{SMOM}}$ 
for a factor of 10 change in gauge-fixing tolerance as a function of $\mu$. 
To cover the residual gauge-fixing effects we take an 
additional uncorrelated uncertainty 
on our $Z_m^{\mathrm{SMOM}}$ results of 0.0004 for $\mu$ = 2 GeV, 0.0001 for $\mu$ = 3 GeV 
and 0.00002 at $\mu$ = 4 GeV on sets 1 through 11. This is typically at the level of our 
statistical uncertainties. On set 12 we fixed to Landau gauge with a tolerance of 
$10^{-14}$ and do not take any additional uncertainty from residual gauge-fixing effects.    
We do not consider any possible effects from Gribov copies (for studies of 
this in the RI-MOM scheme see, for example,~\cite{Giusti:2002rn, Gattringer:2004iv}).

Another way to assess the size of nonperturbative effects is to look 
at differences of $Z$ factors for operators which should have 
the same perturbative expansion. 
Since we are concentrating here on $Z_S$ it makes sense 
to look at the difference between $Z_S$ and $Z_P$. This difference showed 
significant problems with $Z_P$ in the RI-MOM scheme because it exposed 
nonperturbative contributions that behaved as $1/\mu^2$~\cite{Blum:2001sr}. 
This behaviour can be traced to the fact that the inserted operator is 
carrying no momentum ($q=0$) in that scheme~\cite{Martinelli:1994ty}. This causes 
particular problems for the pseudoscalar operator and was deemed to make 
this vertex function of only very limited use. A related issue arises with 
the scalar operator in the RI-MOM scheme, however, and that is one of 
very strong dependence on the quark mass. 
These features were illustrated for the HISQ action in~\cite{Lytle:2015btr} where 
the vertex functions $\Lambda_P$ and $\Lambda_S$ are compared for the 
RI-MOM and RI-SMOM schemes as a function of momentum and quark mass, 
and the superior behaviour of the RI-SMOM scheme is very clear. 

In the RI-SMOM 
scheme, as Figure~\ref{fig:ZPS} shows, the nonperturbative behaviour of 
$\mathrm{Tr}(\Lambda_P-\Lambda_S)/48$ (proportional to $(1/Z_P-1/Z_S)$) 
is quite benign, falling as $1/\mu^6$ with little 
dependence on the lattice spacing. This indicates that 
either $Z_P$ or $Z_S$ could be used to determine the mass renormalisation 
factor; we will however concentrate here on $Z_S$. As we will see in 
Section~\ref{subsec:valmass}, the quark mass dependence of $Z_m$ derived 
from $Z_S$ in 
the RI-SMOM scheme is also much less of an issue than it was in the RI-MOM 
scheme. 

The slope with $1/\mu^6$ of $\mathrm{Tr}(\Lambda_P-\Lambda_S)/48$ in Figure~\ref{fig:ZPS} is 
$\mathcal{O}(1)$ ${\mathrm{GeV}}^6$ when translated into its effect on $Z_P-Z_S$, 
in approximate agreement with what is seen with domain-wall 
quarks~\cite{Aoki:2010dy}. 
This sets an appropriate scale to use when assessing systematic effects 
from nonperturbative contributions in Section~\ref{sec:mass}. These
systematic effects will show up when results evaluated 
at different $\mu$ are run perturbatively in the continuum to a 
common scale. We will use multiple values of $\mu$ (2, 2.5, 3, 4 and 5 GeV) 
in our analysis and compare results for the $\overline{\mathrm{MS}}$ mass at 
a reference scale of 3 GeV.

$Z_m^{\mathrm{SMOM}}$ is dimensionless but the appropriate scale for it, $\mu$, 
must be obtained in GeV by multiplying $a\mu$ by the inverse of the lattice 
spacing. The value of the lattice spacing is obtained from determining 
a dimensionful quantity that can be matched in the continuum limit at 
physical quark masses to an experimental value. We use the Wilson flow 
parameter, $w_0$~\cite{Borsanyi:2012zs}, itself fixed at the value 
0.1715(9) fm using the decay constant of the $\pi$~\cite{fkpi}.  

The physical quark mass limit can be approached in a number of different ways. 
When calculating quantities such as hadron masses, which are sensitive to low 
momentum scales, it is convenient to keep the bare coupling
constant, $\alpha_{\mathrm{lat}}=g^2/(4\pi)$, and $w_0$ fixed. 
This means that the value of $a$ varies slightly with the sea quark masses 
but the variation of hadron masses is small, since they behave in a similar 
way to $w_0$.  An alternative, which is more suitable for the determination of 
bare parameters such as quark masses for reasons discussed in Appendix 
A of~\cite{Chakraborty:2014aca},
is to fix $\alpha_{\mathrm{lat}}$ and the lattice spacing. 
This latter method is the one that we will implement here. 

We use  
results from~\cite{Chakraborty:2014aca} where the 
sea quark mass dependence of $w_0/a$ in terms of the result at the 
tuned physical point is determined for different $\alpha_{\mathrm{lat}}$. 
A universal linear dependence on $\delta m^{\sea}_{uds}$ (the 
deviation of the sum of the $u/d$ and $s$ sea masses 
from their physical values) is 
seen, for values of $\delta m^{\sea}_{uds}$ in 
units of the tuned $s$ quark mass less than 0.5 (see 
Figure 10 in Appendix A). 
An analysis of dependence on the sea $c$ quark mass away 
from the physical point is also given. 
We use the fits of~\cite{Chakraborty:2014aca} to 
interpolate results for $w_0/a$ for sets of ensembles at 
a given value of $\beta$ to the physical quark mass point. 
The values 
that we obtain for $w_0/a$ in this way are given 
for `coarse' ($\beta=6.0$), `fine' ($\beta=6.30$) 
and `superfine' ($\beta=6.72$) sets in Table~\ref{tab:params}. 
We will use 
the $w_0/a$ result (and the value it implies for $a^{-1}$ in GeV) 
for all the ensembles with that value of $\beta$. 

As we will see in Section~\ref{subsec:seamass} this approach 
means that $Z_m^{\mathrm{SMOM}}$ has little discernible sea quark 
mass dependence. This is expected insofar as $Z_m^{\mathrm{SMOM}}$ 
represents physics at the cut-off scale that is a function of 
$\alpha_{\mathrm{lat}}$, with the light sea quark masses having 
only a very small effect on its perturbative expansion. 
We will test the impact of 
the $c$ sea mass in the next section and in Appendix~\ref{appendix:seamass}.    

We take a similar approach for the tuned bare quark masses for $s$ and 
$c$, as will be discussed further in section~\ref{sec:mass}. 

\subsection{Sea mass dependence}
\label{subsec:seamass}

\begin{table}
\caption{Lattice spacing values in units of the Wilson flow 
parameter~\cite{Borsanyi:2012zs}, $w_0$, and tuned quark masses 
for the coarse, fine and superfine sets of ensembles as 
determined in~\cite{Chakraborty:2014aca}. 
These are obtained by fitting the sea quark mass dependence of these parameters 
and interpolating/extrapolating to physical sea quark masses. 
The lattice spacing is obtained from $w_0/a$ by using the value 
for $w_0$ of 0.1715(9) fm determined from the pion decay constant 
in~\cite{fkpi}. For the quark masses the uncertainty is split into 
two pieces. 
The first uncertainty is uncorrelated between lattice spacing values 
and comes from statistical/fitting errors and uncertainties in the 
value of $w_0/a$.  
The second uncertainty is correlated between lattice spacing 
values because it comes from the uncertainty in $w_0$ and from the 
uncertainty in the $\eta_c$ or $\eta_s$ meson mass as appropriate.  
}  
\label{tab:params}
\begin{ruledtabular}
\begin{tabular}{llll}
$\phantom{x}$ &     $w_0/a$ &     $m_c^{\tuned}$ (GeV) &      $m_s^{\tuned}$ (GeV) \\
\hline
       coarse &  1.4075(18) &  1.049(1)(3) &  0.0859(1)(7) \\
         fine &  1.9500(21) &  0.973(1)(3) &  0.0818(1)(7) \\
    superfine &   2.994(10) &  0.901(2)(3) &  0.0768(2)(7) \\
\end{tabular}
\end{ruledtabular}
\end{table}

\begin{figure}[ht]
\includegraphics[width=0.45\textwidth]{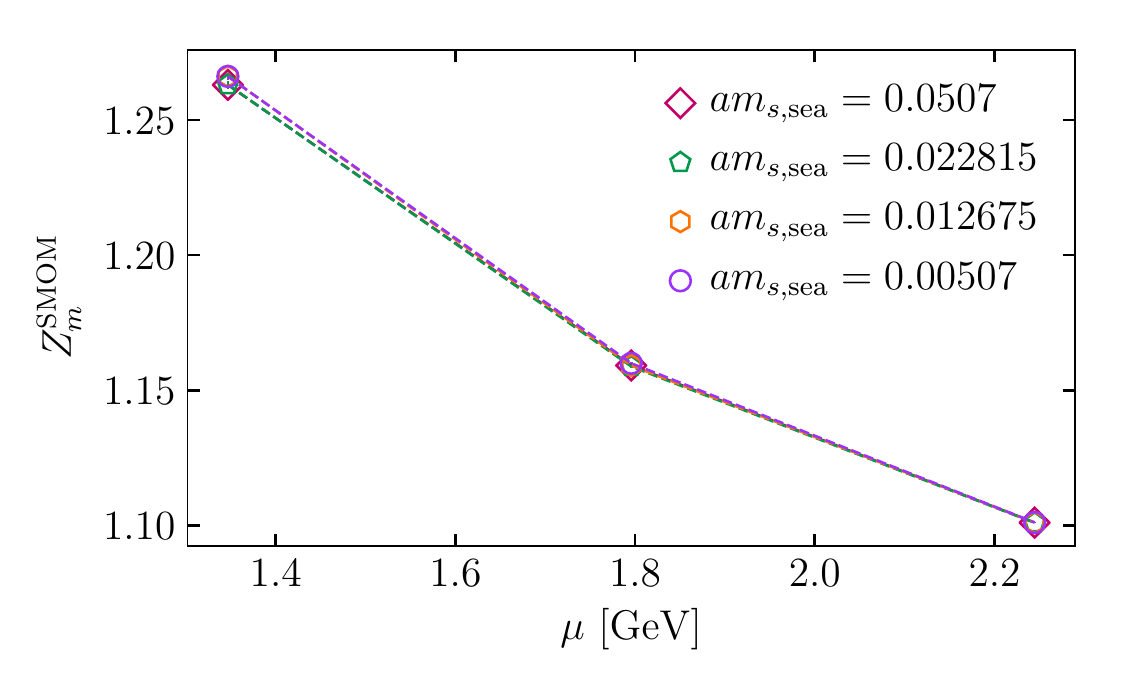}
\includegraphics[width=0.45\textwidth]{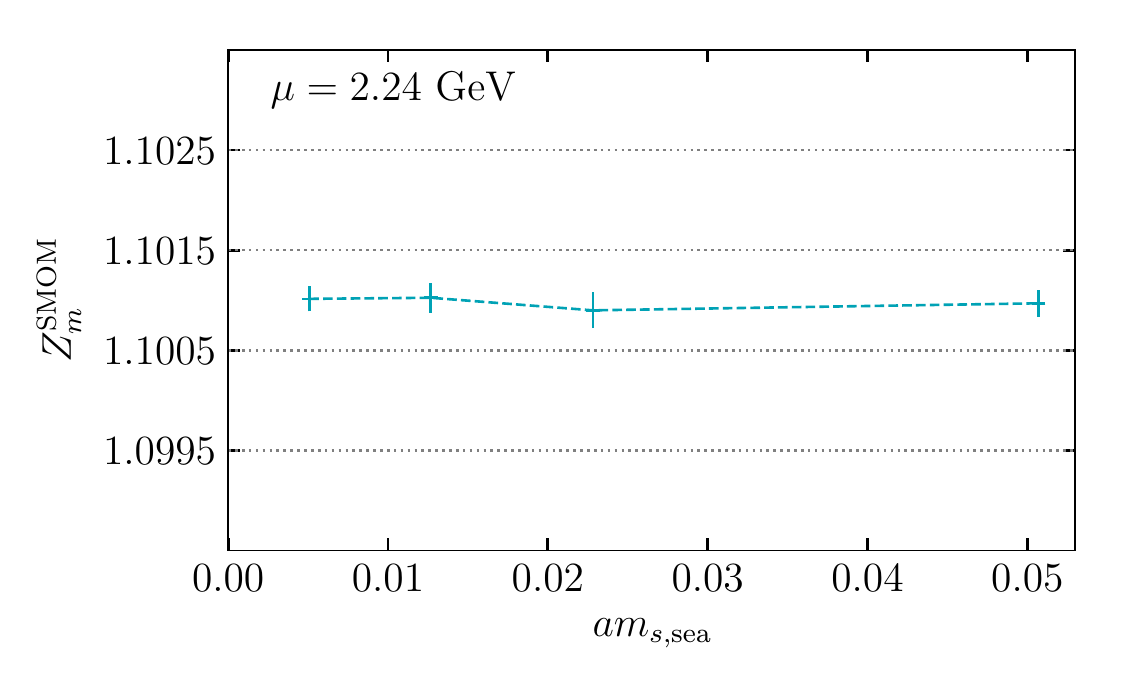}
\caption{
Results for $Z_m^{\mathrm{SMOM}}$ for coarse ensembles with 
different values of the sea $s$ quark mass, with $u/d$ and $c$ 
sea quark masses fixed in lattice units at 0.00507 and 
0.628 respectively (sets 4, 6, 7 and 8, see Table~\ref{tab:ensembles}).
The upper plot shows results for $Z_m^{\mathrm{SMOM}}$ as a function 
of $\mu$ in GeV for the four sets. Dashed lines join the points. 
In all cases the valence quark 
mass is set to 0.0051 in lattice units.  
The lower plot gives more detail for results at $\mu$ = 2.24 GeV, showing 
no visible variation (using the horizontal dotted lines as guides) 
in $Z_m^{\mathrm{SMOM}}$ even at a level below 0.1\%.  
}
\label{fig:mass_sea_s}
\end{figure}

\begin{figure}[ht]
\includegraphics[width=0.45\textwidth]{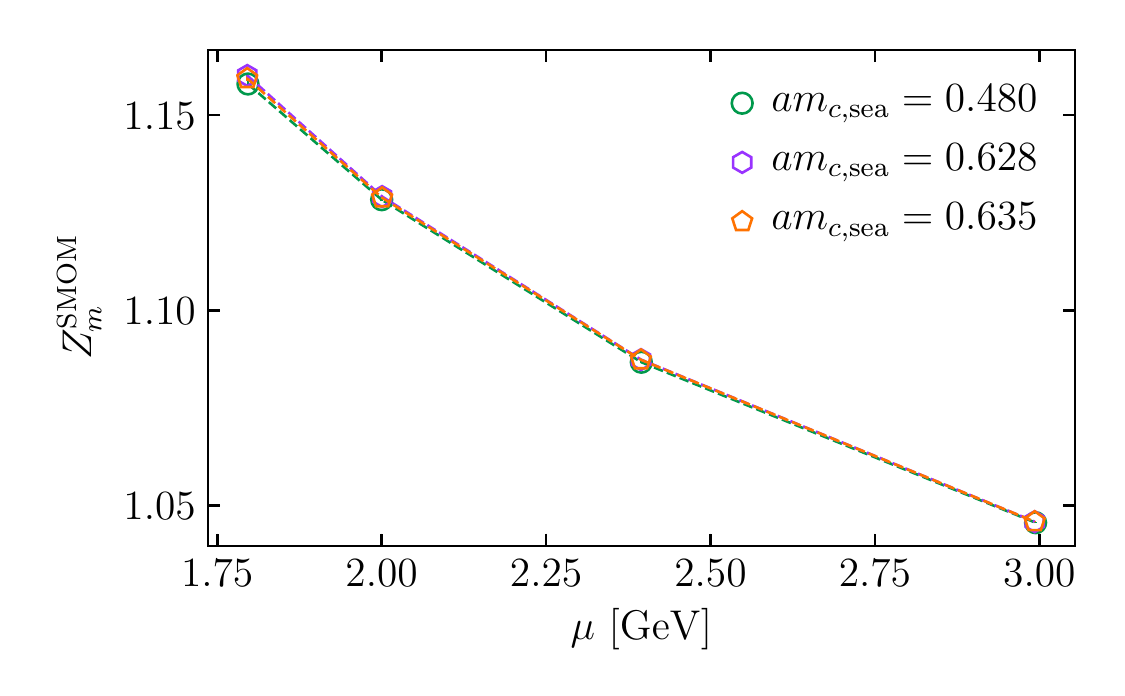}
\includegraphics[width=0.45\textwidth]{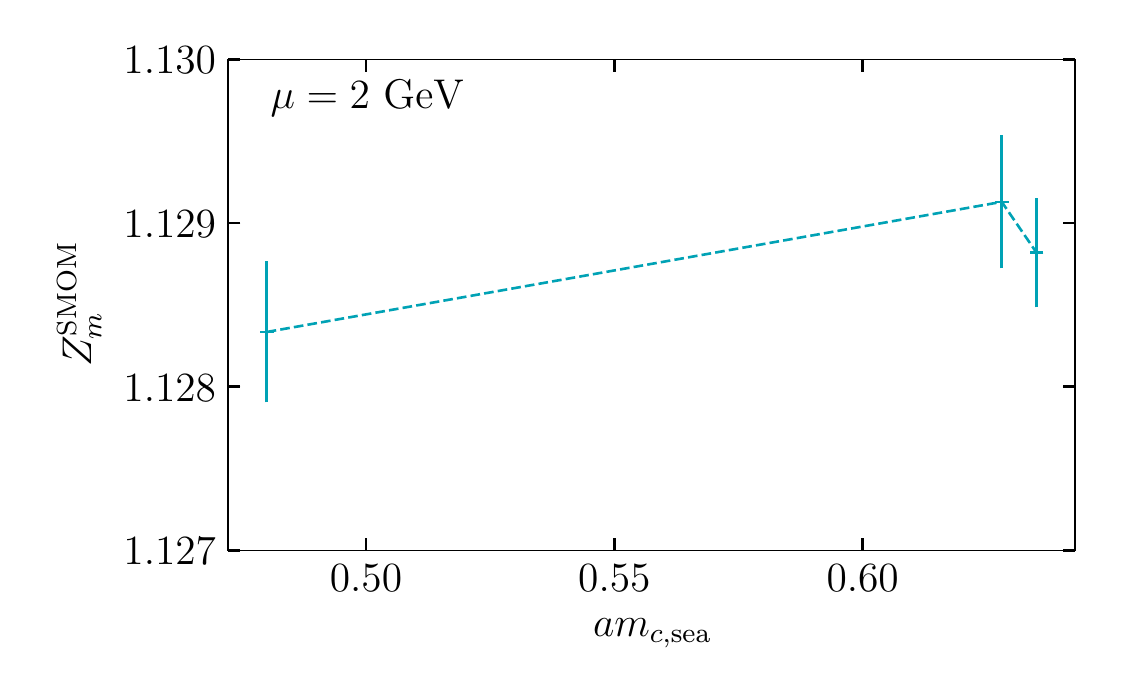}
\caption{
Results for $Z_m^{\mathrm{SMOM}}$ for coarse ensembles with 
different values of the sea $c$ quark mass from sets 1, 2 and 3 
(see Table~\ref{tab:ensembles}). These sets have slightly 
different $u/d$ and $s$ sea quark masses, but by an amount which is 
much smaller than the change in the $c$ sea mass from set 1 to sets 2 and 3. 
The upper plot shows results for $Z_m^{\mathrm{SMOM}}$ as a function 
of $\mu$ in GeV for the two sets. 
The results shown here are obtained for a valence quark mass 
in lattice units of 0.0051. 
The lower plot shows more detail for results at $\mu$=2.4 GeV, showing 
$\mathcal{O}(0.1\%)$ variation for a very substantial change in $am_{c,\sea}$.  
Dashed lines simply join the points. 
}
\label{fig:mass_sea_c}
\end{figure}

\begin{figure}[ht]
\includegraphics[width=0.45\textwidth]{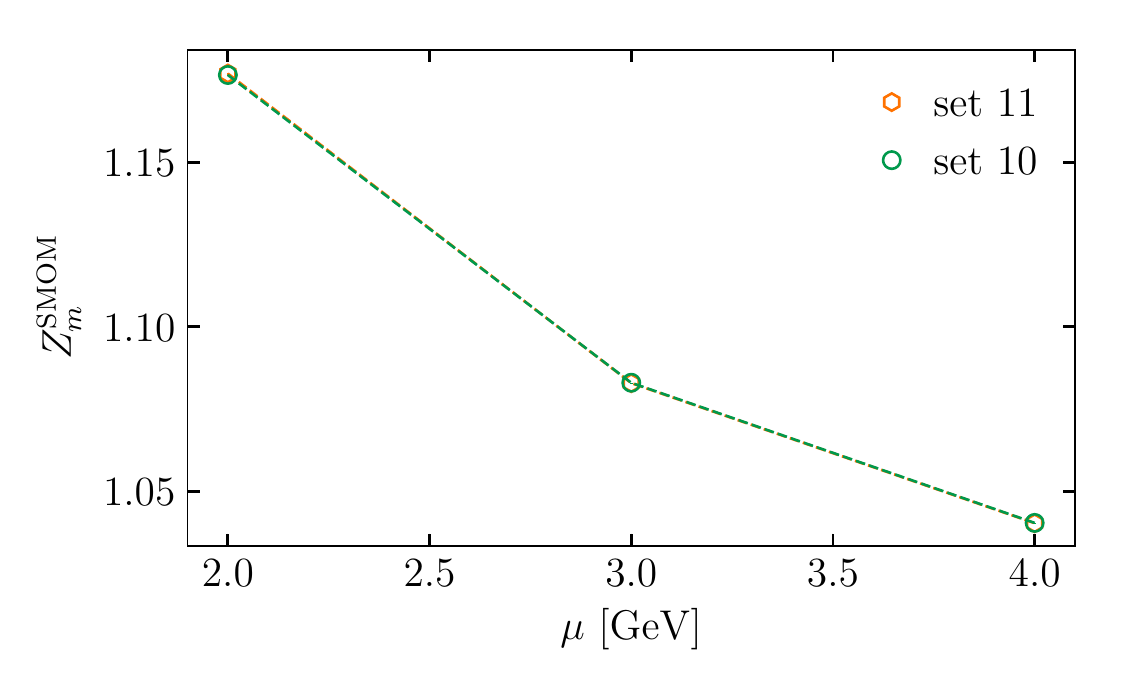}
\caption{
Results for $Z_m^{\mathrm{SMOM}}$ for fine ensembles with 
different values of the sea $u/d$ quark mass from sets 10 and 11
(see Table~\ref{tab:ensembles}). These sets also have slightly 
different $s$ and $c$ sea quark masses and spatial volume. 
The plot shows results for $Z_m^{\mathrm{SMOM}}$ as a function 
of $\mu$ in GeV for the two sets, for a valence mass in lattice 
units of 0.0074. 
}
\label{fig:mass_sea_fine}
\end{figure}

Table~\ref{tab:ensembles} shows the variety of ensembles on which we have 
calculated $Z_m^{\text{SMOM}}$. Note particularly how many different combinations 
of sea quark masses we have studied for $\beta=6.0$ (`coarse'). 
This allows us to test for dependence on the sea quark masses, given the 
method described in Section~\ref{sec:lattice} for fixing the lattice 
spacing. No significant sea quark mass dependence is seen for the 
$\mu$ values that we will use for our analysis. Some of the 
ensembles have very different combinations of sea quark masses from
those that would be considered suitable for a comparison to the real world. 
For example, set 6 has $u/d$ and $s$ quark masses equal at a value around 1/10th 
that of the physical $s$ mass. Nevertheless even this ensemble 
has a $Z_m^{\mathrm{SMOM}}$ 
that agrees (for a given $\mu$) with that from set 4 where $m_s$ is ten times larger 
and therefore more realistic. Note that the components of $Z_m$, i.e. $Z_q$ 
and the vertex function, typically show somewhat larger changes with sea mass 
but the effects cancel in $Z_m$. 

Figure~\ref{fig:mass_sea_s} shows a comparison of $Z_m^{\mathrm{SMOM}}$
for sets 4, 6, 7 and 8 in which $am_s^{\sea}$ varies over a wide range
with no discernible difference, at a level below 0.1\%, in $Z_m^{\mathrm{SMOM}}$, 
for $\mu$ values of 1.8 GeV and above. 
The lowest value of $am_s^{\sea}$ shown in Figure~\ref{fig:mass_sea_s} 
corresponds to the $u/d$ quark mass in the sea. This figure therefore also 
indicates how little variation we can expect as the $u/d$ quark mass in 
the sea is varied. In our final analysis we will include results from 
different values of the sea quark masses to allow for small variations to 
be taken into account, but 
these results indicate that such variations are below the level of our 
statistical uncertainties.  

A similar picture is seen for variation with the $c$ sea mass, 
even though the $c$ sea mass is much larger and $2m_c$ is close to 
$\mu$ in our range of $\mu$ values, so that one could worry that 
an effect might be discernible. We can gauge the possible size 
of such an effect 
from the perturbative analysis of the impact of massive 
$c$ quarks in the sea given in Appendix~\ref{appendix:seamass}.
That shows a shift of size 0.1\% for a change in $m_c$ from zero to $m_c$ 
at $\mu$ = 2 GeV. 
Figure~\ref{fig:mass_sea_c} shows a comparison of results for 
$Z_m^{\mathrm{SMOM}}$ as a function of $\mu$ for sets 1 and 2 
which have a substantially (30\%) different $c$ sea quark mass, along 
with slightly different $u/d$ and $s$ sea quark masses (which 
Figure~\ref{fig:mass_sea_s} has already demonstrated have 
no effect) and different spatial volumes (again for which we see no effect 
in Section~\ref{subsec:volume}).
 We also include results for set 3 which has a value of 
$am_c$ differing from that on set 2 by 1\%, a size of variation which is closer 
to that of typical $m_c$ mistuning on our ensembles~\cite{Chakraborty:2014aca}.  
The lower plot of Figure~\ref{fig:mass_sea_c} gives more detail at 
$\mu$ = 2 GeV and shows, as expected, no variation 
of $Z_m^{\text{SMOM}}$ at the level of 0.1\% for 
a change in $am_c$ of 30\%.  It also shows that there is no impact on our 
results at the level of our statistical errors from the slight (5\%) mistuning 
of the $c$ sea mass that we have on some ensembles. 

Finally, in Figure~\ref{fig:mass_sea_fine} we compare results for two fine 
lattices with different sea mass values (sets 10 and 11).  
This plot covers three $\mu$ values we will use in our final analysis, 
2, 3 and 4 GeV. Good agreement is seen between the results on sets 10 
and 11 (with the largest discrepancy being 0.1\% for $\mu$ = 2 GeV), testing 
sea-mass dependence as well as dependence on the spatial volume, 
to be discussed in more detail in Section~\ref{subsec:volume}.  

\subsection{Valence mass dependence \& extrapolation}
\label{subsec:valmass}

\begin{figure}[ht]
\includegraphics[width=0.45\textwidth]{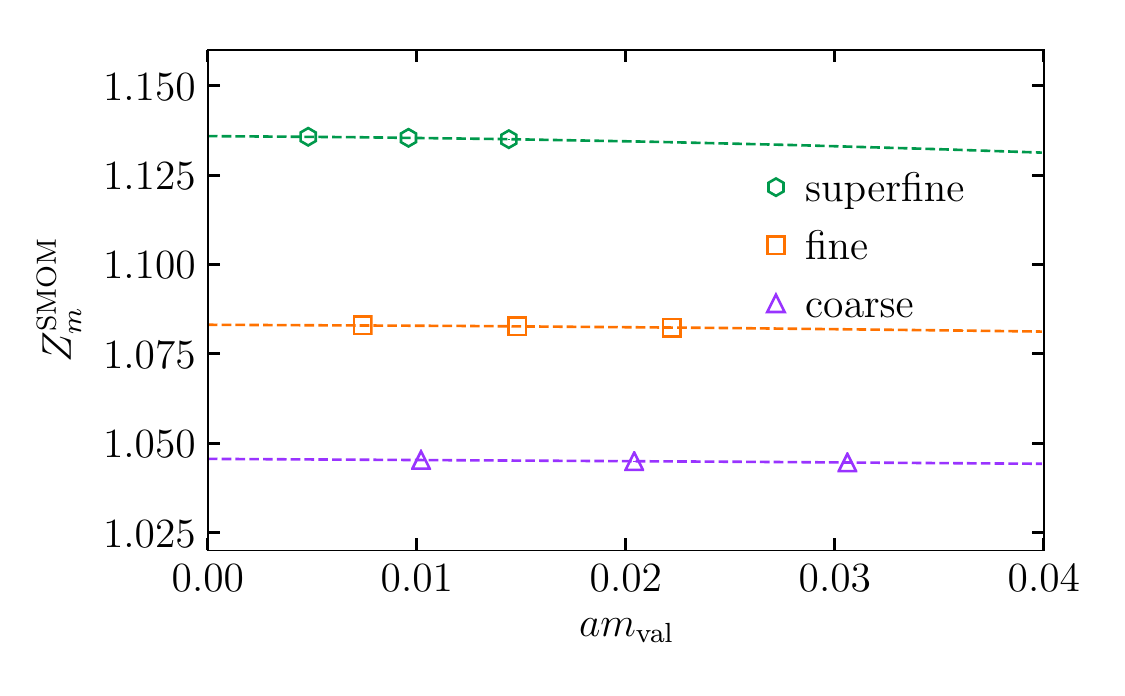}
\includegraphics[width=0.45\textwidth]{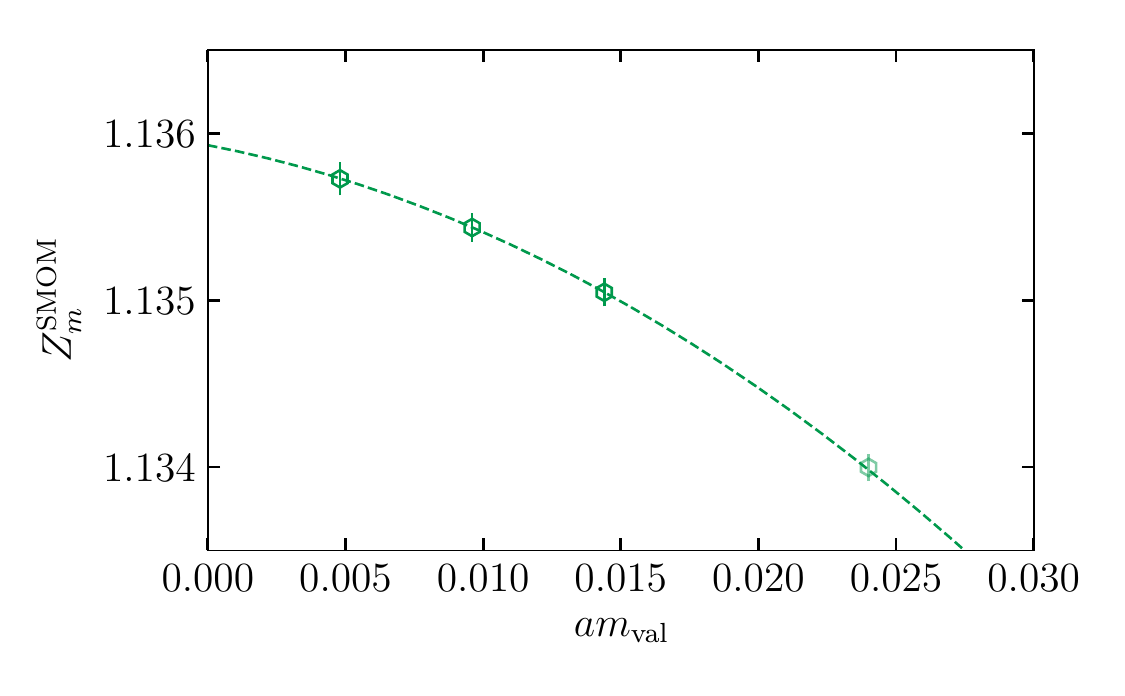}
\caption{ The upper plot shows the 
dependence of $Z^{\text{SMOM}}_m$ on valence quark mass in lattice units 
for coarse, fine and superfine lattices (sets 2, 10 and 12) at $\mu$ = 3 GeV. 
The dashed line gives the simple fit described in the text.  
The lower plot is a zoomed in version of the superfine (set 12) results 
for which fit parameters are shown in Figure~\ref{fig:valslope}. 
The lighter rightmost point corresponds to a quark mass equal to 
that of the strange quark. This point was not included in the valence 
quark mass fit, but lies on top of the fitted line. 
}
\label{fig:valmass}
\end{figure}

\begin{figure}[ht]
\includegraphics[width=0.45\textwidth]{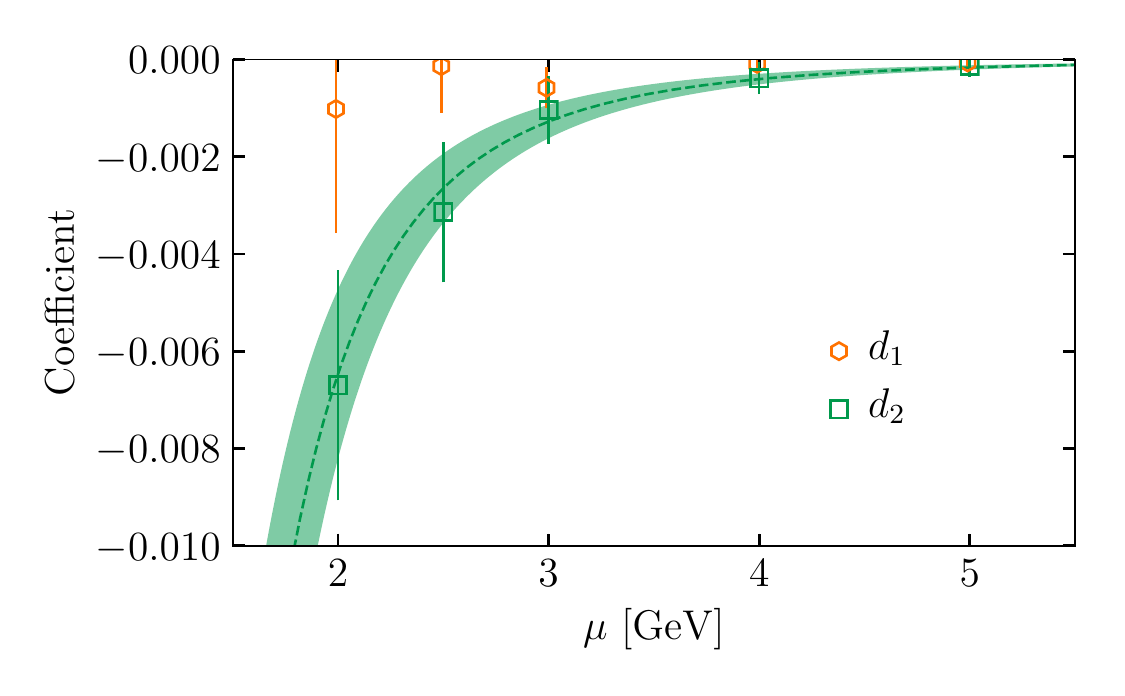}
\caption{ The linear and quadratic slopes ($d_1$ and $d_2$ of eq.~(\ref{eq:fitval})) of 
$Z^{\text{SMOM}}_m$ with valence quark mass on superfine lattices (set 12),  
plotted against $\mu$ in GeV. 
The curve with error band is a fit to the form $C/\mu^4$ through the 
$d_2$ points. 
}
\label{fig:valslope}
\end{figure}

In Section~\ref{subsec:seamass} the $Z_m^{\mathrm{SMOM}}$ renormalisation factors 
are determined for small and fixed but non-zero valence quark masses
and we showed that the dependence on sea quark masses is almost negligible. 
Here we will show that there is a small but visible dependence for $Z_m^{\mathrm{SMOM}}$ 
on the valence quark mass. 
This dependence comes 
from the vertex function since the wave function renormalisation is almost 
independent of the valence quark mass. 
Since the impact in perturbation theory of the small valence quark masses we 
use should be negligible, the most likely source of valence quark 
mass dependence 
is nonperturbative, i.e. that of quark masses multiplying a condensate 
appearing in conjunction with inverse powers of $\mu$ 
as expected from the OPE. 

Figure~\ref{fig:valmass} shows the dependence of $Z^{\mathrm{SMOM}}_m$ 
on valence mass in lattice units, $ma$, for a coarse, fine and superfine ensemble 
(sets 2, 10 and 12)
at a fixed value of $\mu$ (3 GeV). 
For each case we determine $Z_m^{\mathrm{SMOM}}$ for three valence masses; 
that of the $u/d$ quark mass in the sea and two and three times that value. 
Figure~\ref{fig:valmass} shows very little visible dependence on $ma$ 
but it is, however, 
significant (see the lower plot of Figure~\ref{fig:valmass} for 
more detail in the superfine case). Note that the results at different values of $am$ are 
correlated and we include this correlation in our fits through a 
covariance matrix determined by the bootstrap procedure.

As discussed in Section~\ref{sec:lattice} it is convenient to extrapolate 
in the valence quark mass to zero, so that we can convert from the $\mathrm{SMOM}$ 
scheme to the $\overline{\mathrm{MS}}$ scheme using perturbation theory at 
zero valence quark mass. The extrapolation also has the advantage of removing some of 
the non-perturbative condensate contributions that are not part of $Z^{\mathrm{SMOM}}_m$. 
The size of the observed mass dependence then provides an indication of the size 
of the remaining condensate contributions that do not depend on the quark mass. 

To extrapolate to zero valence quark mass we fit $Z_m^{\mathrm{SMOM}}$ 
to a simple polynomial form in $am_{\mathrm{val}}$ given below. Including only a 
linear term does not give a good fit over the range of $am$ values that we use 
when correlations are included. We therefore add in both a quadratic and cubic term: 
\begin{eqnarray}
\label{eq:fitval}
Z_m^{\mathrm{SMOM}}(am_{\mathrm{val}}) &=& Z_m^{\mathrm{SMOM}} + d_1\frac{am_{\mathrm{val}}}{am_s} + \\
&& d_2 \left(\frac{am_{\mathrm{val}}}{am_s}\right)^2 + d_3 \left(\frac{am_{\mathrm{val}}}{am_s}\right)^3. \nonumber
\end{eqnarray}
We use a prior 
on $Z_m^{\mathrm{SMOM}}$ (the value of $Z_m^{\mathrm{SMOM}}(am)$ in the massless 
limit) of $1.0 \pm 0.5$. For the coefficients $d_i$ priors of $\{0 \pm 0.1,0 \pm 0.01,0 \pm 0.001\}$ are used
for $\mu = 2\ \text{GeV}$ with the values being decreased by a factor of 2 and 4, respectively 
for $\mu = 3\ \text{GeV}$ and 4 GeV to allow for an approximate $\mu^{-2}$ suppression, 
the smallest inverse power of $\mu$ that we expect to appear.  
We divide the lattice valence masses by the tuned $s$ quark mass at that 
lattice spacing so that the $d_i$ are dimensionless and physical. 
Note that, if the linear term were set by the gauge-invariant 
condensate $m\langle \overline{\psi}\psi \rangle/\mu^4$, we would expect $d_1$ to 
be $\mathcal{O}(2\times 10^{-4})$, which would make the mass dependence 
scarcely visible. 
Instead our priors allow for possibly larger 
gauge-noninvariant condensates to appear.   
The linear slope shows significant lattice spacing dependence and is consistent 
with zero on the superfine lattices, as shown in Figure~\ref{fig:valslope}. 
The coefficient of the quadratic mass dependence, $d_2$, is non-zero and is shown 
for the superfine lattices in Figure~\ref{fig:valslope} along with the results 
of a simple fit to the form $C/\mu^4$ ($C/\mu^2$ does not give a good fit, although $C/\mu^6$ is 
also acceptable) 
with $C$ = -0.10(3) $\mathrm{GeV}^4$, equivalent to $-(0.56(4)\,\mathrm{GeV})^4$. 

From this (and earlier results in Section~\ref{sec:lattice}) we conclude that 
in our fits in Section~\ref{sec:mass}, 
comparing quark masses determined using $Z_m^{\mathrm{SMOM}}$ 
values at different values of $\mu$, we should allow for condensate contributions 
remaining in $Z_m^{\mathrm{SMOM}}$ that could be as large 
as $(1 \mathrm{GeV})^n/\mu^n$ coming from gauge-noninvariant 
condensates. This will allow us to include an uncertainty from such 
nonperturbative contributions in our determination of the mass. 

\subsection{Volume dependence}
\label{subsec:volume}

\begin{figure}[ht]
\includegraphics[width=0.45\textwidth]{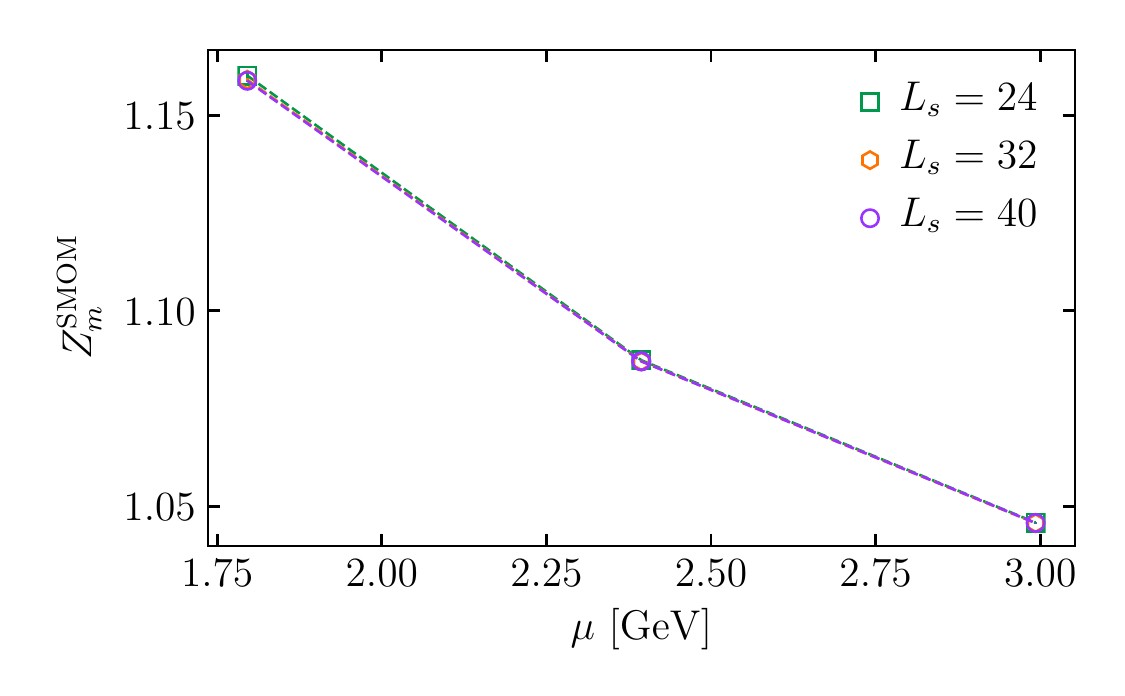}
\includegraphics[width=0.45\textwidth]{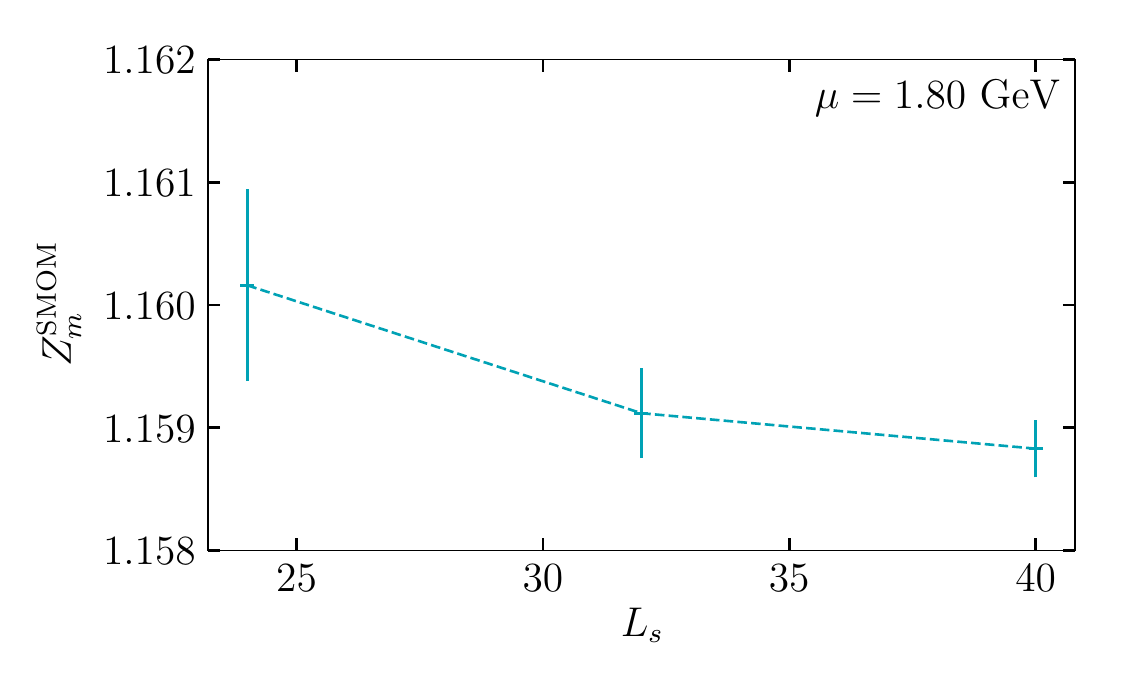}
\caption{ Dependence of $Z^{\text{SMOM}}_m$ on the spatial volume of the lattice 
(sets 3, 4 and 5) as a function of $\mu$.  
The valence quark mass in lattice units is fixed to 0.0051. 
}
\label{fig:volume}
\end{figure}

Since $Z_m^{\mathrm{SMOM}}$ is a matching factor between two different 
regularisations of QCD we expect it to be dominated by ultraviolet physics 
and not to be sensitive to the volume of the lattice. However, we have observed 
some infrared sensitivity in the form of nonperturbative condensate 
contributions. In aiming for a precise 
determination of $Z_m^{\mathrm{SMOM}}$ finite-volume effects need to be tested. 
This is straightforward to do 
on lattices that have the same $\beta$ and sea quark 
masses and differ only by the number of lattice points in each spatial direction. 
Figure~\ref{fig:volume} shows such results for sets 3, 4 and 5 that have 24, 32 and 40 
lattice points in each spatial direction but exactly the same parameters in the 
lattice QCD Lagrangian (see Table~\ref{tab:ensembles}). 
No significant dependence on the lattice size is seen except for 
very small $\mu$ (below 2.0 GeV, which is our smallest value for analysis) 
and for small lattices (of size $L_s$=24 which is smaller in terms of 
$M_{\pi}L_s$ than any of the lattices that we use for analysis). 
We conclude from this that finite volume effects are negligible.

\section{Determination of masses in the $\overline{\text{MS}}$ scheme}
\label{sec:mass}

\begin{table*}
\caption{ Results for $Z_m^{\mathrm{SMOM}}$ for 
$\mu=$ 2, 2.5, 3, 4 and 5 GeV on a subset of ensembles from 
Table~\ref{tab:ensembles} covering coarse to superfine 
lattice spacings. The values are obtained by extrapolating 
to zero valence quark mass as discussed in the text. 
For each set of results we also give, in column 5, the correlation matrix 
between the values for different $\mu$. 
Note that there is very slight mistuning of some $\mu$ values on 
the coarse and fine lattices (sets 2 -- 11) and the actual $\mu$ values are given 
in the column headings. 
Statistical errors only are given here. We include a further uncorrelated 
uncertainty on $Z_m$ values for sets 2 -- 11 as described 
in Section~\ref{sec:lattice} to account for residual 
gauge-fixing effects ($\pm 0.0004$ at $\mu$ = 2 GeV, $\pm 0.0001$ at 3 GeV and 
$\pm 0.00002$ at 4 GeV).  
}  
\label{tab:ZmSMOM}
\begin{ruledtabular}
\begin{tabular}{lllllll}
\multicolumn{7}{c} {$Z_m^{\mathrm{SMOM}}(\mu)$, $\mu$ in GeV : } \\
\hline
Set & 2.004 & 2.500 & 3.005 & 4.007  &  & correlation\\
\hline 
2 & 1.12967(40) & 1.07935(20) & 1.045628(90) & - & - & $\left(\begin{array}{lll} 1 & 0.41 & 0.12 \\ 0.41 & 1 & 0.45 \\ 0.12 & 0.45 & 1 \end{array}\right)$ \\
4 & 1.12990(42) & - & 1.045434(61) & - & - & $\left(\begin{array}{ll} 1 & -0.17 \\ -0.17 & 1 \end{array}\right)$ \\
9 & 1.13061(22) & - & 1.045518(53) & - & - & $\left(\begin{array}{ll} 1 & 0.33 \\ 0.33 & 1 \end{array}\right)$ \\
10 & 1.17726(45) & 1.11954(15) & 1.083082(77) & 1.040445(25) & - & $\left(\begin{array}{llll} 1 & -0.19 & 0.41 & 0.52 \\ -0.19 & 1 & -0.21 & -0.13 \\ 0.41 & -0.21 & 1 & 0.42 \\ 0.52 & -0.13 & 0.42 & 1 \end{array}\right)$ \\
11 & 1.17748(35) & - & 1.082955(55) & 1.040350(23) & - & $\left(\begin{array}{lll} 1 & 0.16 & 0.36 \\ 0.16 & 1 & 0.72 \\ 0.36 & 0.72 & 1 \end{array}\right)$ \\
\hline
 & 2.000 & 2.500 & 3.000 & 4.000  & 5.000 & correlation\\
\hline 
12 & 1.24884(93) & 1.18100(31) & 1.13662(12) & 1.083481(54) & 1.053782(32) & $\left(\begin{array}{lllll} 1 & 0.35 & 0.26 & 0.19 & 0.22 \\ 0.35 & 1 & 0.32 & 0.45 & 0.22 \\ 0.26 & 0.32 & 1 & 0.26 & 0.10 \\ 0.19 & 0.45 & 0.26 & 1 & 0.59 \\ 0.22 & 0.22 & 0.10 & 0.59 & 1 \end{array}\right)$ \\
\end{tabular}
\end{ruledtabular}
\end{table*}

\begin{table}
\caption{ Results for $Z_m^{\overline{\mathrm{MS}}/\text{SMOM}}$ for 
$\mu=$ 2, 2.5, 3, 4 and 5 GeV using eq.~(\ref{eq:Zmpert}) and 
values from Tables~\ref{tab:mom-msbar} and~\ref{tab:deltaCm}. 
The uncertainty in $Z$ quoted here comes from the uncertainty in the value of 
$\alpha_s$ (and so is 100\% correlated between the values). We use 
$\alpha_{\overline{\mathrm{MS}}}(n_f=4,5.0\,\mathrm{GeV})$ = 0.2128(25)~\cite{Chakraborty:2014aca}. 
Column 3 gives the multiplicative factor $R(3\,\text{GeV}, \mu)$ that
converts the $\text{MS}$ mass at scale $\mu$ to the mass at the 
reference scale of 3 GeV. This is obtained from 
4-loop running in perturbative QCD. The 
uncertainty is dominated by that in $\alpha_s$; 
the uncertainty from missing higher order terms in the running is negligible 
in comparison. The error in $R$ is then 100\% 
correlated or anti-correlated between the values, depending on whether 
$R$ is greater than or less than 1. There is also a 100\% correlation (or anticorrelation, 
as appropriate) with the errors in $Z_m^{\overline{\mathrm{MS}}/\text{SMOM}}$. 
Note that the values in the table are for the $\mu$ values in column 1. We 
calculate $R$ and $Z$ inside our fit function and hence allow for the fact 
there that the $\mu$ values are slightly mistuned on coarse 
and fine lattices (see Table~\ref{tab:ZmSMOM}). 
We include an additional uncertainty, as described in the 
text, to allow for $w_0/a$ errors feeding in to the determination of $\mu$. 
This gives a (correlated) uncertainty of 0.0003 on the coarse lattices, 0.0002 
on the fine lattices and 0.0008 on the superfine lattices.   
}  
\label{tab:Zmpert}
\begin{ruledtabular}
\begin{tabular}{lll}
$\mu$~(GeV) & $Z_m^{\overline{\mathrm{MS}}/\text{SMOM}}(\mu)$ & $R(3\,\text{GeV}, \mu)$\\
\hline
 2.0 & 0.9792(5) & 0.9034(20) \\
 2.5 & 0.9821(3)          & 0.9582(8)          \\
 3.0 & 0.9838(3) & -  \\
 4.0 & 0.9859(2) & 1.0616(11) \\
 5.0 & 0.9871(2) & 1.1063(19) \\
\end{tabular}
\end{ruledtabular}
\end{table}

\begin{figure}[ht]
\includegraphics[width=0.45\textwidth]{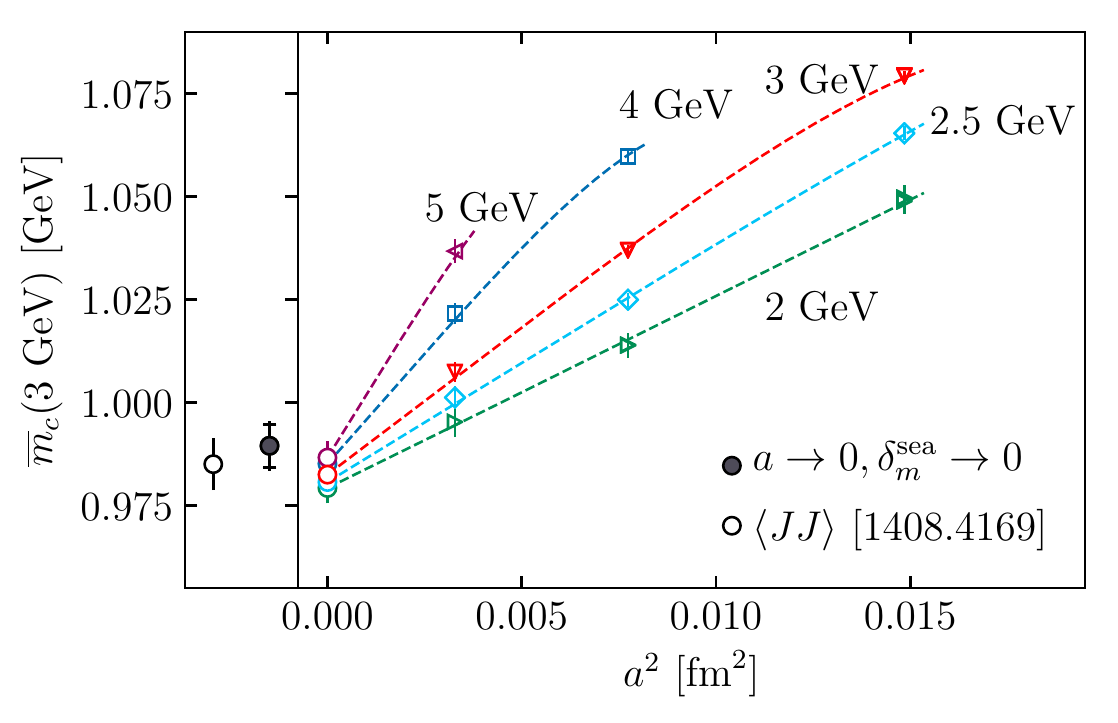}
\includegraphics[width=0.45\textwidth]{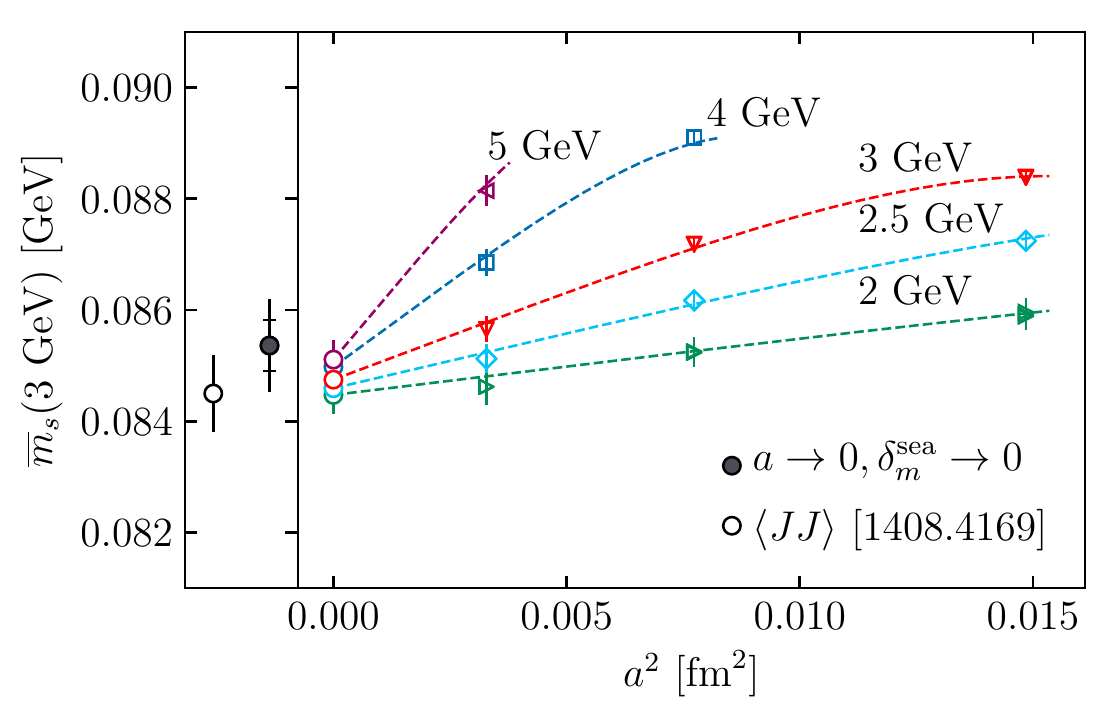}
\caption{
(Upper) $\overline{m}_{c}(3 \GeV)$ and 
(lower) $\overline{m}_{s}(3 \GeV)$ obtained from nonperturbative 
lattice QCD calculations of $Z_m^{\mathrm{SMOM}}$ 
at three different values of $\mu$, followed by perturbative matching 
to the $\overline{\mathrm{MS}}$ scheme and running to 3 GeV. 
The results are plotted for $\mu$ = 2, 2.5, 3, 4 and 5 GeV against the square 
of the lattice spacing. Extrapolations to the continuum limit for each value 
of $\mu$, as discussed in the text, are shown as dashed lines
(these give the fit result at $\delta_{\ell}^{\sea}=\delta_c^{\sea} =0$).   
The error bars on the data points 
show only the uncorrelated uncertainties.
The point plotted as a dark grey filled circle, offset to the left, is 
the final physical result, $\overline{m}$(3 GeV),
from the fit described in the text. The inner error bar for this point is 
the uncorrelated uncertainty and the outer error bar the full uncertainty.  
The point plotted as a white circle, further offset to the left, is from 
the current-current correlator method~\cite{Chakraborty:2014aca} (along with a 
nonperturbative determination of $m_c/m_s$ in the $m_s$ case). 
}
\label{fig:mass_extrap}
\end{figure}

Our procedure here for determining the quark mass in the $\overline{\mathrm{MS}}$ 
scheme has three ingredients: 
\begin{itemize}
\item A quark mass in our lattice QCD scheme tuned nonperturbatively to 
reproduce the mass of a given hadron from experiment;
\item A nonperturbative determination from lattice QCD of the mass 
renormalisation factor $Z_m^{\mathrm{SMOM}}$ 
that converts this mass at each value of the lattice spacing 
to a mass in our SMOM scheme at 
a given value of the scale, $\mu$;
\item  A perturbative calculation of the mass renormalisation factor 
$Z_m^{\overline{\mathrm{MS}}/\text{SMOM}}$ (through $\alpha_s^2$) 
that further converts 
the SMOM mass at scale $\mu$ to the mass in the $\overline{\mathrm{MS}}$ 
scheme at scale $\mu$. From there we can run the mass to different 
scales using 4-loop running in the $\overline{\mathrm{MS}}$ 
scheme~\cite{Chetyrkin:1997dh, Vermaseren:1997fq}. 
\end{itemize}
Then 
\begin{equation}
\label{eq:masscalc}
\overline{m}(\mu,a) = Z_m^{\overline{\mathrm{MS}}/\text{SMOM}}(\alpha_s(\mu))Z_m^{\mathrm{SMOM}}(\mu,a)m(a) .
\end{equation}
Here $m(a)$ is the bare lattice quark mass, in physical units, at a specific 
value of the lattice spacing, the first item from the list above. 
$Z_m^{\mathrm{SMOM}}$ is the mass renormalisation factor calculated nonperturbatively 
on lattice QCD configurations at a specific lattice spacing, allowing us to convert 
the lattice mass to the SMOM scheme at a scale $\mu$. How this is calculated 
has been discussed in earlier sections; here we will give the results. The intermediate 
quark mass we obtain in the SMOM scheme, although nominally now in a continuum 
scheme at a physical scale, will still carry remnants of its lattice origins through 
discretisation errors. These must be removed by calculation at multiple values of 
the lattice spacing, so that an extrapolation to the continuum limit, $a=0$, can be made. 
This could be done with the SMOM masses, but we choose to first convert to 
the $\overline{\mathrm{MS}}$ scheme at scale $\mu$ by multiplying by the final factor 
$Z_m^{\overline{\mathrm{MS}}/\text{SMOM}}$. We denote the $\overline{\mathrm{MS}}$ 
mass obtained this way as $\overline{m}(\mu,a)$ to show that it has yet 
to be extrapolated to the continuum limit. We will describe how we do this below; first 
we give the results that we will use for each of the ingredients of eq.~(\ref{eq:masscalc}). 

\begin{table}
\caption{ Error budget, giving a breakdown of the uncertainties 
in the $c$ and $s$ quark masses in the $\overline{\mathrm{MS}}$ scheme at 
a scale of 3 GeV obtained from the fits described in the text. All 
the uncertainties are given as a percentage of the final answer. 
The condensate uncertainties include all the uncertainties 
from that term in the fit function, which also allows for discretisation 
and $m_{\sea}$ effects. 
}
\label{tab:errbudget}
\begin{ruledtabular}
\begin{tabular}{rll}
$\phantom{x}$ & $\overline{m}_{c}$(3 GeV) & $\overline{m}_{s}$( 3 GeV)\\
\hline
$a^{2} \to 0$ & 0.28 & 0.28 \\
Missing $\alpha_{s}^{3}$ term & 0.22 & 0.22  \\
Condensate & 0.23 & 0.23 \\
$m_{\sea}$ effects & 0.00 & 0.00 \\
$Z^{\overline{\mathrm{MS}}/\mathrm{SMOM}}_m$ and $R$ & 0.04 & 0.04 \\
$Z^{\mathrm{SMOM}}_{m}$ & 0.13 & 0.13 \\
Uncorrelated $m^{\tuned}$ & 0.20 & 0.23 \\
Correlated $m^{\tuned}$ & 0.30 & 0.82 \\
Gauge fixing & 0.11 & 0.11 \\
$\mu$ error from $w_0$ & 0.12 & 0.12 \\
\hline
Total: & 0.62\% & 0.99\% \\
\end{tabular}
\end{ruledtabular}
\end{table}

\subsection{$m(a)$}
\label{subsec:latticemass}
The bare lattice quark masses that we use are for $s$ and $c$ quarks and are 
given in~\cite{Chakraborty:2014aca} for the ensembles and lattice spacing values 
that we use here. The $c$ quark mass, $m_c(a)$, was tuned by adjusting the lattice mass 
to give the physical value for the $\eta_c$ meson mass. The physical value for 
the $\eta_c$ mass was defined from the experimental value with a shift upwards 
by 2.7 MeV (less than 
0.1\%) to remove electromagnetic effects 
and to account for $c\overline{c}$ annihilation, since both of these effects are 
missing in our lattice QCD calculation~\cite{Davies:2010ip, Gregory:2010gm}. 
The uncertainty on the physical $\eta_c$ mass 
to which we match is then increased (to 2.7 MeV) 
to allow for uncertainty in these corrections. 
The $s$ quark mass is similarly tuned based on the physical mass of the 
$s\overline{s}$ pseudoscalar particle known as the $\eta_s$. It is an unphysical 
particle since its valence quarks are artificially not allowed to annihilate, but 
its properties can be well determined in lattice QCD~\cite{Davies:2009tsa, fkpi}.
Its mass can be determined in terms of $K$ and $\pi$ meson masses as~\cite{fkpi}
\begin{equation}
\label{eq:etasmass}
M_{\eta_s}^{\mathrm{phys}} = 0.6885(22)\,\mathrm{GeV} .
\end{equation}
where the uncertainty includes a systematic error from the neglect of electromagnetism 
in the lattice QCD calculations.  

The tuned lattice bare $c$ and $s$ quark masses are given in GeV 
in Table II of~\cite{Chakraborty:2014aca}. These are the values that 
give the physical $\eta_c$ or $\eta_s$ mass on each ensemble given 
a fixed value for $w_0$. Since here (as explained in Section~\ref{sec:lattice}) 
we are approaching the physical mass point using a fixed lattice spacing value 
(since that removes sea quark mass dependence from our results, as shown 
in Section~\ref{subsec:seamass}) then we also need fixed tuned quark mass for 
sets of ensembles at a fixed value of $\beta$. The fits to the dependence on sea quark 
mass discussed in Appendix A of~\cite{Chakraborty:2014aca} enables us to 
determine the tuned $c$ and $s$ quark masses for physical sea quark masses 
at each value of $\beta$. These are the values that we will use here 
and they are given in Table~\ref{tab:params}. 

The uncertainties in the tuned masses include the uncertainty from the 
lattice spacing. This gives a $3\times$ smaller relative uncertainty for 
the $c$ quark mass than for the $s$ quark mass because the lattice spacing 
uncertainty appears with the `binding energy' of the meson rather than its mass. 
For the $\eta_c$ the binding energy is much smaller than the mass, but for 
the $\eta_s$ it is of the same size.  
Table~\ref{tab:params} divides the uncertainty on the tuned masses into 
two components: a correlated uncertainty from the value of $w_0$ and the 
value of the meson masses used to tune the quark mass that is 
the same for all lattice spacing values, and an uncertainty that is 
uncorrelated between lattice spacing values since it comes, for example, from 
statistics/fitting or the values of $w_0/a$ determined separately for each 
ensemble. 

\subsection{$Z_m^{\mathrm{SMOM}}$}
\label{subsec:ZmSMOM}
Working from right to left in eq.~(\ref{eq:masscalc})
the next set of results that we need are for $Z_m^{\mathrm{SMOM}}$
for each ensemble that we will use in determining our continuum and chiral 
limit for the quark masses. We have chosen to work with multiple
values of $\mu$ in order to assess the impact of nonperturbative 
terms on the mass renormalisation. These are $\mu$ = 2, 2.5, 3, 4 and 5 GeV.  
At each value of $\mu$ we determine $Z_m^{\mathrm{SMOM}}$ at three 
values of the valence quark mass, as described in Section~\ref{subsec:valmass}, 
and extrapolate to zero valence mass. Results are given in Table~\ref{tab:ZmSMOM} 
for the ensembles that we will use. The results for different $\mu$ 
values on a given ensemble are correlated and so we include in the 
Table the correlation matrix for the $Z$ values. 
The correlation matrix, $\rho_{ij}$, for variable $x_i$ and $x_j$ is defined by 
\begin{equation}
\label{eq:cov}
\rho_{ij} = \frac{\langle x_{i}x_{j} \rangle - \langle x_{i}\rangle \langle x_{j}\rangle}{\sigma_i\sigma_j}  
\end{equation}
with $\langle \rangle $ indicating the expectation 
value, and $\sigma$ the standard deviation.

Results are given for sets 2, 4, 9, 10, 11 and 12 that we will use 
in our final analysis which will determine a continuum limit for the 
mass and allow for small residual sea quark mass effects. 
Note the very slight mistunings of $\mu$ from the nominal values 
on sets 2--11. These are allowed for in our fits. 

\subsection{$Z_m^{\overline{\mathrm{MS}}/\mathrm{SMOM}}$}
\label{subsec:MSBARSMOM}
The third ingredient for eq.~(\ref{eq:masscalc}) is the 
matching coefficient from SMOM to $\overline{\text{MS}}$. 
For this we use the perturbative expansion of eq.~(\ref{eq:Zmpert}) 
with $c_1$ and $c_2$ values from Table~\ref{tab:mom-msbar} 
for the RI-SMOM case and for $n_f=4$. Values for $\alpha_s$ in 
the $\overline{\text{MS}}$ scheme at the different values of 
$\mu$ are given in Table~\ref{tab:deltaCm}. 
We use the results of Appendix~\ref{appendix:seamass} to adjust 
$c_2$ to allow for having a massive $c$ quark in the sea. This has a 
very small effect for $\mu=$ 2 GeV, and even smaller one for 
$\mu=$ 2.5 GeV and is otherwise negligible. 
The resulting values for $Z_m^{\overline{\mathrm{MS}}/\text{SMOM}}$ 
are given in Table~\ref{tab:Zmpert}. 
The uncertainty in the $Z$ values quoted there comes from the uncertainty 
in $\alpha_s$ and so is 100\% correlated between the values. 
There is also a systematic uncertainty from missing 
higher orders in the perturbative expansion and we will allow for 
that in our final fits below.

\subsection{Fitting $\overline{m}(\mu,a)$ to determine $\overline{m}(3\,\mathrm{GeV})$}
\label{subsec:running}
By multiplying all three ingredients together as in eq.~(\ref{eq:masscalc}) 
we obtain values for the $c$ or $s$ quark mass in the $\overline{\text{MS}}$ 
scheme at a (nominal) scale of $\mu$ = 2, 2.5, 3, 4 or 5 GeV from each lattice ensemble.  
These results still contain discretisation effects from the lattice QCD 
component of the calculation. To remove these effects we must
extrapolate to the continuum limit. At the same time we want to assess 
other systematic effects such as the nonperturbative contributions to 
the lattice QCD determination of $Z_m^{\mathrm{SMOM}}$ that have 
not been removed by our extrapolation 
to zero valence quark mass, and missing higher order perturbative 
contributions to $Z_m^{\overline{\mathrm{MS}}/\text{SMOM}}$. 
This can be done by comparing results at different $\mu$ but, 
the simplest way to pick out these effects is 
to run all the results to a common scale, $\mu_{\text{ref}}$. 
We take $\mu_{\text{ref}}$ = 3 GeV, so that we run up from 2 and 2.5 GeV and 
down from 4 and 5 GeV. The running is done by integrating the 
evolution equations numerically in the $\overline{\text{MS}}$ scheme 
using 4-loop 
expressions~\cite{vanRitbergen:1997va, Chetyrkin:1997dh, Vermaseren:1997fq, Czakon:2004bu}.   
The result of this is a multiplicative factor $R(\mu_{\text{ref}},\mu)$ such 
that 
\begin{equation}
\label{eq:run}
\overline{m}(\mu_{\text{ref}}) = R(\mu_{\text{ref}},\mu)\overline{m}(\mu) .
\end{equation}
Values of $R(3\,\text{GeV},\mu)$ are given in Table~\ref{tab:Zmpert}. 
Note that, because the uncertainty comes from the uncertainty in $\alpha_s$ 
the uncertainty is 100\% (anti-)correlated 
between the different $\mu$ values. The uncertainties in $R$ and  
$Z_m^{\overline{\mathrm{MS}}/\text{SMOM}}$ are also 100\% (anti-)correlated 
for the same reason. 

At this point we also need to include an uncertainty coming from the 
relative determination of the lattice spacing on coarse, fine and 
superfine lattices. The uncertainty in $w_0/a$, given in Table~\ref{tab:params},
means that the $\mu$ values on each set may not match and this gives 
an additional uncertainty in the running of the mass to the 
3 GeV reference point, including in the values obtained at $\mu$ = 3 GeV. 
This gives an additional (correlated) uncertainty of 0.0003 on the coarse 
lattices, 0.0002 on fine lattices and 0.0008 on superfine lattices. 
There is an additional correlated 0.1\% uncertainty on all points 
coming from the effect on $\mu$ of the uncertainty in the value of $w_0$. 

We then have results for $\overline{m}_s(\mu_{\text{ref}})$ and 
$\overline{m}_c(\mu_{\text{ref}})$ that come from lattice 
calculations with different values of the lattice spacing and different 
values of $\mu$. We fit these to a function that allows for discretisation 
effects that depend on $a$ and other systematic effects that depend 
on $\mu$. It is important to include the correlations between the 
points: our results at different values of $a$ are correlated through 
their dependence on the value of $w_0$ which is used to determine 
the lattice spacing and our results at different $\mu$ for a given ensemble 
are correlated through the statistical uncertainties in the values of 
$Z_m^{\overline{\mathrm{MS}}/\mathrm{SMOM}}$ (see Table~\ref{tab:ZmSMOM}). 

Our results are plotted in Figure~\ref{fig:mass_extrap}. 
Discretisation effects are clearly evident with the slope in $a^2$ becoming 
larger with larger $\mu$, not surprisingly. Results at different $\mu$ 
come together on the finer lattices. 

A key point, as we have emphasised, is to provide constraints on nonperturbative 
$\mu$ dependence (from condensate terms) that would survive the continuum limit 
from our lattice QCD calculation
but is not part of $Z_m$. To understand how big these terms might be, we turn 
to the OPE (for more details, see Appendix~\ref{appendix:ope}). The analysis of the quark propagator is 
given in~\cite{Lavelle:1988eg, Lavelle:1992yh, Chetyrkin:2009kh}. 
To lowest order in inverse powers of the momentum ($p \equiv \mu$) 
and $\alpha_s$ this gives (rather than eq.~(\ref{eq:Zq})) 
\begin{equation}
\label{eq:Zqcond}
\frac{1}{12p^2} \Tr[S^{-1}(p) \, \Xsl{p}] = -Z_q + \frac{\pi \alpha_s(p)}{3}\frac{\langle A^2 \rangle}{p^2} + \mathcal{O}(X/p^4) .
\end{equation}
Here $Z_q$ is the perturbative contribution from the leading (unit) operator, 
$A^2$ is the square of the Landau gauge gluon field 
and $X$ denotes vacuum expectation 
values of dimension 4 operators such as $m\overline{\psi}\psi$ (which vanishes 
at zero quark mass), $\overline{\psi}A\psi$ and $G^2$.  
The coefficients of the higher dimension operators in the OPE are obtained by 
matching scattering amplitudes for both sides of the OPE from, 
for example, low-momentum gluons.  
Repeating this procedure for the scalar vertex function for the symmetric kinematic 
point that we use in this calculation (and which allows an OPE treatment), yields
\begin{equation}
\label{eq:ZScond}
\frac{1}{12} \Tr[\Lambda_S(p_1,p_2)]|_\sym = \frac{Z_q}{Z_S} - \frac{\pi \alpha_s(\mu)}{3}\frac{\langle A^2 \rangle}{{\mu}^2} + \mathcal{O}(X/{\mu}^4) 
\end{equation}
rather than eq.~(\ref{eq:ZSdef}). 
From eqs~(\ref{eq:Zqcond}) and~(\ref{eq:ZScond}) we see that the leading 
nonperturbative contribution to our determination of $Z_m$ is $-(2\pi/3)\alpha_s(\mu)\langle A^2\rangle/\mu^2$.
The value of $\langle A^2 \rangle$ is not well-known~\cite{Blossier:2010ky, Burger:2012ti}
and so we simply allow for it to be of size $\mathcal{O}(1 \mathrm{GeV})^2$ in our fits 
to obtain the continuum limit of the $\overline{\text{MS}}$ quark mass. 
We must also allow for the higher dimension condensates denoted $X$ above, about which 
even less is known.  To do this we include terms at $1/\mu^4$ and $1/\mu^6$ in our 
fits, again allowing the operator vacuum expectation values (summed over all 
the operators that could appear) to be $\mathcal{O}(1 \mathrm{GeV})^{2n}$. 

In fitting our results to obtain physical values for the quark masses we must 
then allow for: lattice spacing artefacts, non-perturbative effects, sea quark mass 
effects and missing terms in the perturbative matching to $\overline{\text{MS}}$. 
To allow for all of these, we fit our results to the following form: 
\begin{eqnarray}
\label{eq:massfit}
\overline{m}(\mu_{\mathrm{ref}},\mu,a) &=& \overline{m}(\mu_{\mathrm{ref}})\times \\
&& \hspace{-6.0em} \left[ 1+ \sum_{n=1}^4 c^{(n)}_{\Lambda^2a^2}(\Lambda a/\pi)^{2n} \right] \times\nonumber \\
&&  \hspace{-5.0em} \bigg( 1 + \sum_{n=1}^{10}c^{(n)}_{\mu^2a^2}(\mu a/\pi)^{2n} + 
   c_{\alpha} \alpha^3_{\MSbar}(\mu) + \nonumber \\
&& \hspace{-4.0em} h_{\ell}^{\sea} \frac{\delta_{\ell}^{\sea}}{m_{s}^{\tuned}} + h_{c}^{\sea}
\frac{\delta_{c}^{\sea}}{m_{c}^{\tuned}} + \nonumber \\
&& \left [ 1 + k_{\ell}^{\sea} \frac{\delta_{\ell}^{\sea}}{m_{s}^{\tuned}} +
k_{c}^{\sea} \frac{\delta_{c}^{\sea}}{m_{c}^{\tuned}} \right ] \times  \nonumber \\ 
&& \hspace{-6.0em} \sum_{n=1}^3 c^{(n)}_{\mathrm{cond}}\alpha_{\MSbar}(\mu) \frac{(1\,\mathrm{GeV})^{2n}}{\mu^{2n}} \times \left[ 1+ c^{(n)}_{\mathrm{cond},a^2}(\tilde{\Lambda} a/\pi)^2 \right] \bigg) . \nonumber 
\end{eqnarray}
Here $\overline{m}(\mu_{\mathrm{ref}})$ is the physical result. 
The coefficients $c_{\mu^2a^2}$ allow for discretisation effects 
set by the scale $\mu$ (coming from $Z_m$) and the $c_{\Lambda^2a^2}$ allow 
for those set by the scale $\Lambda$ in the tuning of the 
bare quark masses, independent of $\mu$.  
We take $\Lambda$ to be 500 MeV in the case of the $s$ quark mass,
 but 1 GeV in the  
$c$ quark mass case, since it could be set by $m_c$ itself. 
We take the prior on all the $c_{a^2}$ coefficients to be $0.0\pm 1.0$. 
$c_{\alpha}$ allows for systematic uncertainties in the continuum limit from missing $\alpha_s^3$ 
terms in the matching of SMOM to $\overline{\text{MS}}$. We take the prior on $c_{\alpha}$ to 
be $0.0\pm 0.2$, allowing for a size four times larger than $c_1$ or $c_2$. 
The coefficients $c_{\mathrm{cond}}$ allow for 
nonperturbative condensate effects that have not been removed by extrapolating the 
valence quark masses to zero. We include three such terms with inverse powers of 
$\mu$ of 2, 4 and 6 since we expect these to be the most significant. 
The results 
that we give in Sections~\ref{sec:lattice} and~\ref{subsec:valmass} 
show that we need to allow for gauge-noninvariant 
condensates of size as large as $(1 \,\mathrm{GeV})^{2n}$. 
We take this to be the generic size of the condensate and 
give each one a coefficient with prior $0\pm 2$ (consistent with 
the combination of eq.~(\ref{eq:Zqcond}) and~(\ref{eq:ZScond})).
We also allow for $a$-dependence in each condensate with $\tilde{\Lambda}$ = 500 MeV 
and a prior on $c_{\mathrm{cond},a^2}$ of $0.0 \pm 1.0$. 
All $h^{\sea}$ and $k^{\sea}$ coefficients allow for any small remaining 
dependence on the sea quark masses, either explicitly in condensate terms 
or elsewhere with
\begin{eqnarray}
\delta_l^{\mathrm{sea}} &=&  \sum_{q=u,d,s} (m_q - m_q^{\mathrm{tuned}})    \nonumber \\
\delta_c^{\mathrm{sea}} &=& m_c - m_c^{\mathrm{tuned}} .      
\end{eqnarray}
We take the priors on these coefficients to be $0.0 \pm 0.2$ 
(consistent with results in Section~\ref{subsec:seamass}). 

The fit is strongly constrained by the number of different correlations included 
between results from different $\mu$ values and different $a$ values. 
We obtain a $\chi^2/\mathrm{dof}$ of 0.8 
for both the fits for $m_c$ and for $m_s$.  
We can also do both fits simultaneously, requiring all coefficients to 
be the same except those for the $(a\Lambda)^n$ terms and then we obtain 
a $\chi^2/\mathrm{dof}$ of 0.7. 
If we drop the condensate terms from the separate fits the $\chi^2/\mathrm{dof}$ 
increases to 2, indicating that these are important. 
The $\chi^2/\mathrm{dof}$ for the simultaneous fit without condensates increases 
to 6. 
We find the total condensate contribution at $a=0$ and $\delta^{\mathrm{sea}}=0$ 
to be relatively small, at -1.0(5)\% at $\mu$ = 2 GeV and -0.3(1)\% at 
$\mu$ = 5 GeV for the separate fits. The simultaneous fit has somewhat more significance, 
at -1.4(4)\% for the condensate contribution at $\mu$ = 2 GeV. 
The condensate contribution is a combination of a negative term at 
$1/\mu^2$ (as expected from above), a positive term at 
$1/\mu^4$ and a relatively unconstrained term at $1/\mu^6$. 

\begin{figure}
\includegraphics[width=0.47\textwidth]{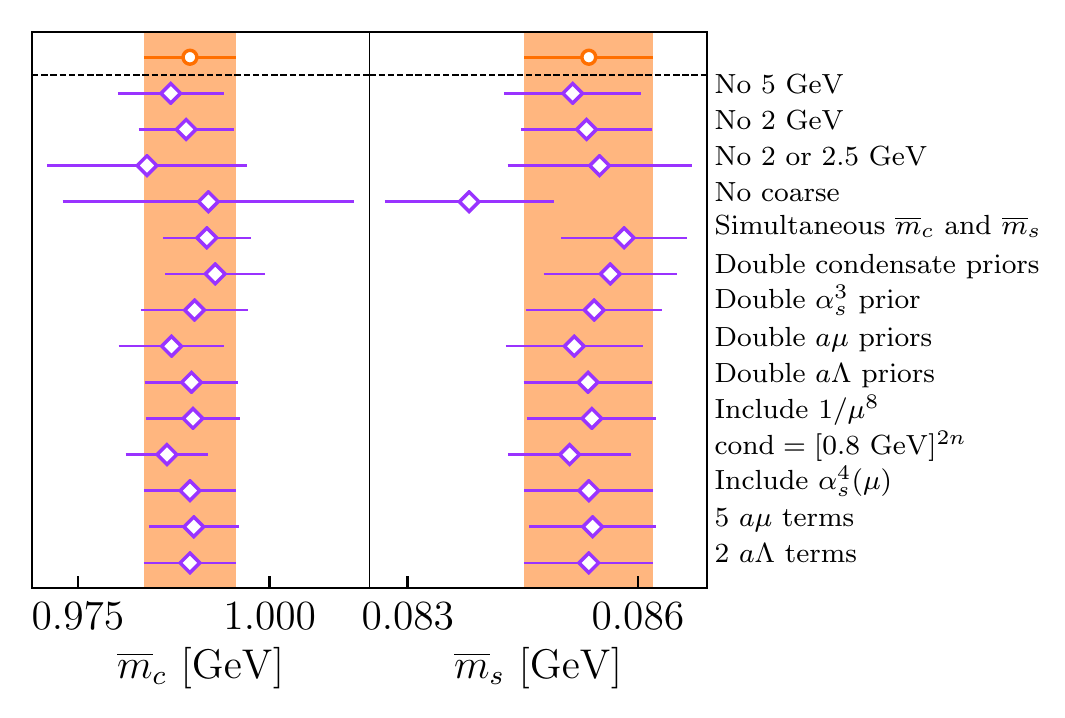}
\caption{ A graphical representation of the different tests that we 
have done to check the robustness of our fits to obtain the physical 
results for $\overline{m}_s$ and $\overline{m}_c$. 
The different rows give variations on the fit described in the 
text (eq.~(\ref{eq:massfit})).  
    } 
\label{fig:robustness}
\end{figure}

Figure~\ref{fig:robustness} demonstrates the robustness of our fit, by showing 
the impact on the final value of numerous modifications. 
These include missing out sets of results; doubling prior widths on various fit 
coefficients and changing the numbers of terms used to describe discretisation effects, 
condensate contributions or missing pieces of the perturbative matching. 
Effects are relatively minor and generally well within our uncertainties. 

The values we obtain for the physical result, $\overline{m}(\mu_{\mathrm{ref}})$ at the 
reference scale of 3 GeV (using separate fits to each mass) are: 
\begin{eqnarray}
\label{eq:m3valsres}
\overline{m}_c (3 \GeV, n_f=4) &=& 0.9896(61)\, \mathrm{GeV} \\
\overline{m}_s (3 \GeV, n_f=4) &=& 0.08536(85)\, \mathrm{GeV} \nonumber
\end{eqnarray}
The error budget for the two numbers, evaluated from the fit, is given in 
Table~\ref{tab:errbudget}. The uncertainties are dominated by those from the 
tuned bare quark masses (especially for $m_s$) but with sizeable contributions 
from the continuum extrapolation, possible 
missing $\alpha_s^3$ terms in the $\mathrm{SMOM}$ 
to $\overline{\mathrm{MS}}$ matching and condensate effects. 

\section{Conclusions}
\label{sec:conclusions}

\begin{figure}
\includegraphics[width=0.45\textwidth]{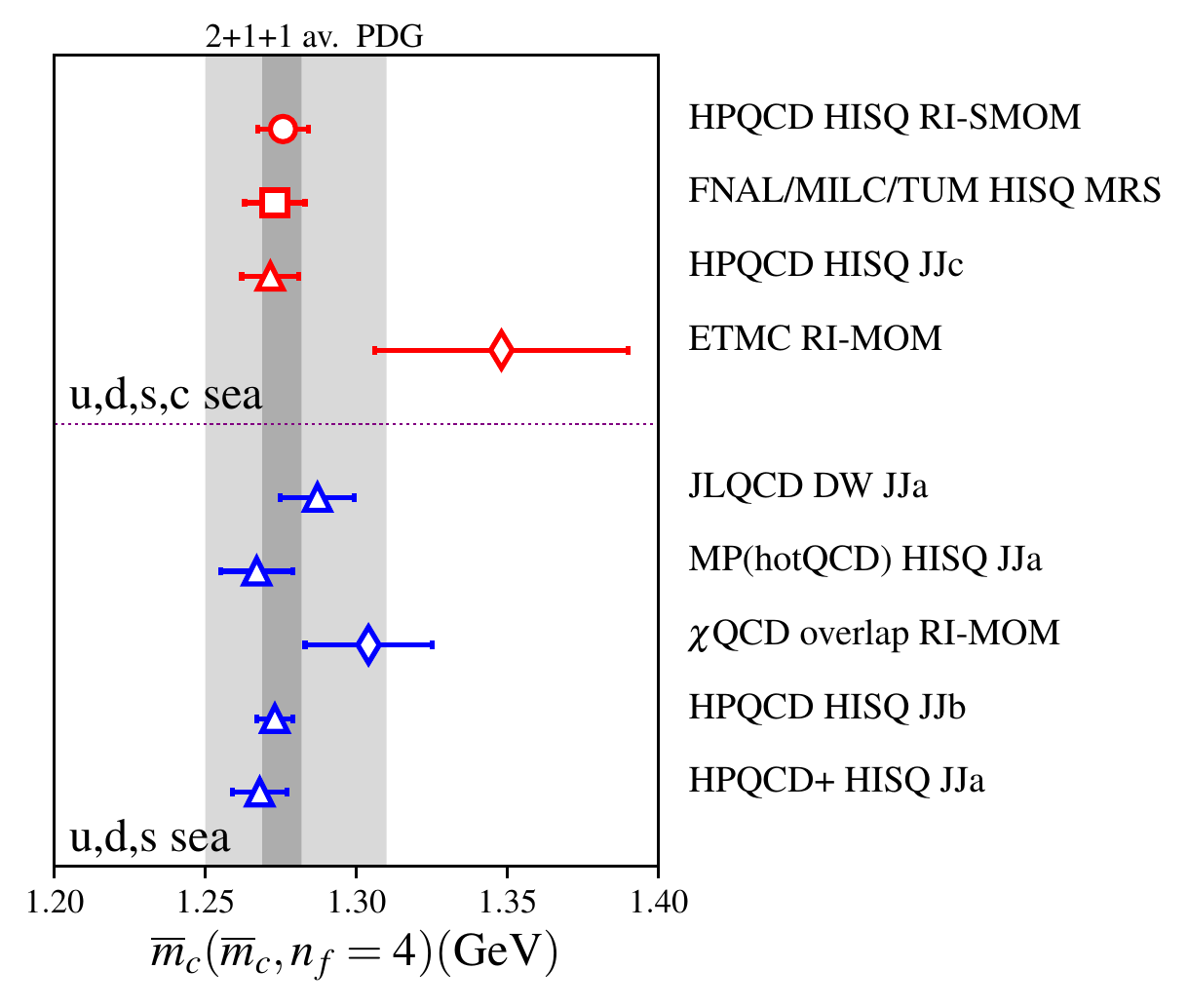}
\caption{ Comparison of lattice QCD results 
for $\overline{m}_c(\overline{m}_c, n_f=4)$. Note that results are 
determined at a higher energy scale and then run down to $\overline{m}_c$ using 
perturbation theory. Calculations are listed chronologically and 
divided into those obtained on gluon configurations that include 
3 (blue symbol) or 4 (red symbol) flavours of quarks in the sea. 
Results obtained on 3 flavour configurations have in all cases been 
adjusted to $n_f=4$ using perturbation theory (see the references for 
details of how this is done in each case). 
Different symbols denote different methods: open triangles, the current-current 
correlator (JJ) method (`a',`b' and `c' variants, see text); open diamonds use the 
RI-MOM intermediate scheme, open circles, the RI-SMOM intermediate scheme and 
open squares, the MRS scheme. 
Labels on the right indicate collaboration name, quark formalism 
and method. 
The result denoted `HPQCD HISQ RI-SMOM' is this work, 
`FNAL/MILC/TUM HISQ MRS' is from~\cite{Bazavov:2018omf}, 
`HPQCD HISQ JJ' (red) is from~\cite{Chakraborty:2014aca}, 
`ETMC RI-MOM' is from~\cite{Carrasco:2014cwa}, 
`JLQCD DW JJ' is from~\cite{Nakayama:2016atf}, 
`MP(hotQCD) HISQ JJ' is from~\cite{Maezawa:2016vgv}, 
`$\chi$QCD overlap RI-MOM' is from~\cite{Yang:2014sea},
`HPQCD HISQ JJ'(blue) is from~\cite{McNeile:2010ji} and 
`HPQCD+ HISQ JJ' is from~\cite{Allison:2008xk}. The grey shaded band 
indicates the world average (taking correlations into account) of 
the $n_f=4$ results and the lighter shaded band
shows the evaluation of $1.28(3)$ GeV from the Particle Data Group~\cite{pdg}.    } 
\label{fig:mccomp}
\end{figure}

\begin{figure}
\includegraphics[width=0.45\textwidth]{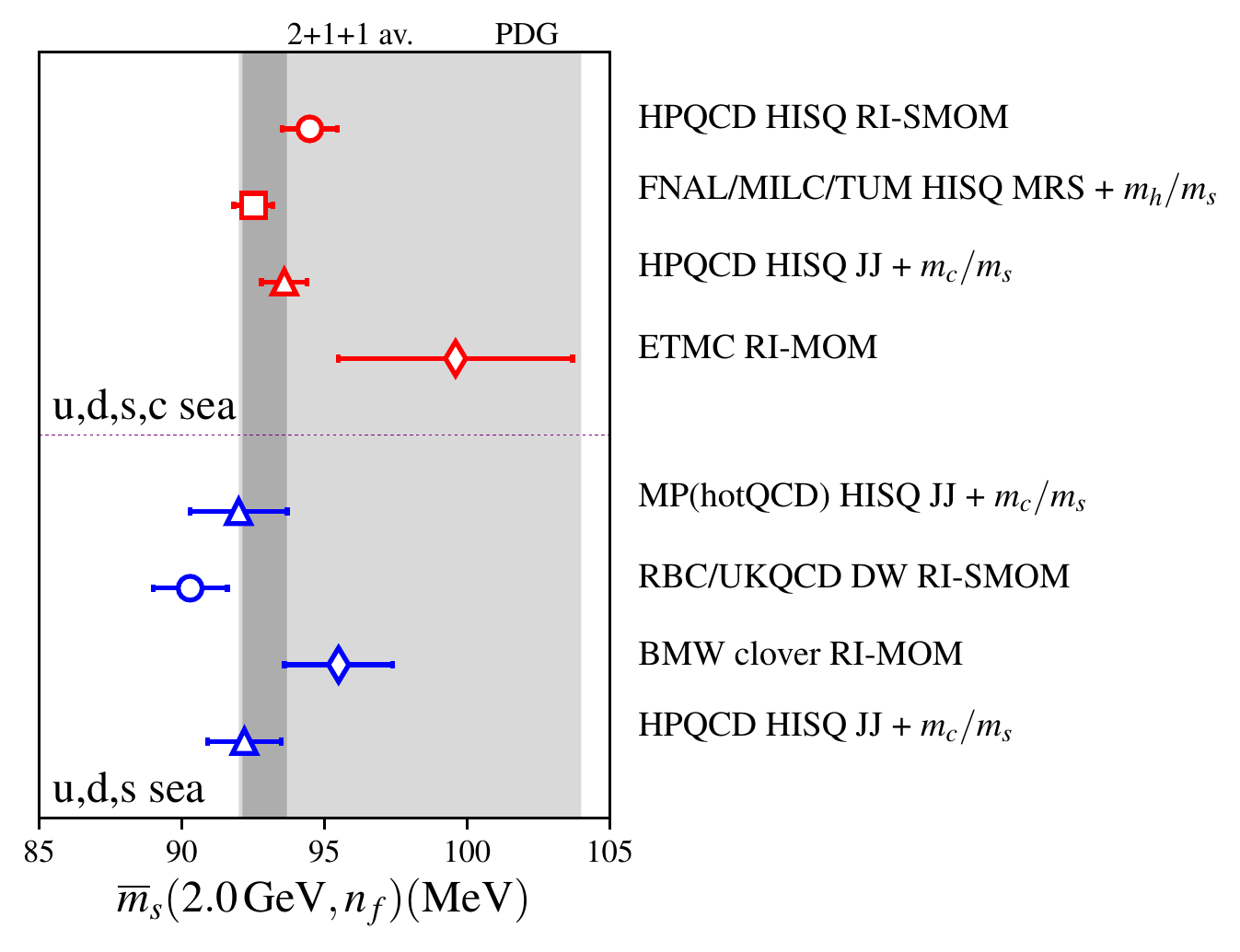}
\caption[jkl]{ Comparison of lattice QCD results 
for $\overline{m}_s(2.0\,\mathrm{GeV}, n_f)$. 
Results are listed chronologically and 
divided into those obtained on gluon configurations that include 
$n_f=3$ (blue symbol) or $n_f=4$ (red symbol) flavours of quarks in the sea. 
The $n_f=4$ results should be 0.2 MeV smaller that $n_f=3$ from 
QCD perturbation theory matching for adding/removing a $c$ sea quark. 
Different symbols denote different methods: open triangles, the current-current 
correlator (JJ) method combined with a nonperturbative determination 
of $m_c/m_s$; open diamonds use the 
RI-MOM intermediate scheme; open circles, the RI-SMOM intermediate scheme 
and open squares the MRS scheme combined with a nonperturbative determination 
of a mass ratio between heavy and $s$ quarks. 
Labels on the right indicate collaboration name, quark formalism 
and method. 
The result denoted `HPQCD HISQ RI-SMOM' is this work, 
`FNAL/MILC/TUM HISQ MRS + $m_h/m_s$' is from~\cite{Bazavov:2018omf},
`HPQCD HISQ JJ + $m_c/m_s$' (red) is from~\cite{Chakraborty:2014aca},
`ETMC RI-MOM' is from~\cite{Carrasco:2014cwa}, 
`MP(hotQCD) HISQ JJ + $m_c/m_s$' is from~\cite{Maezawa:2016vgv}, 
`RBC/UKQCD DW RI-SMOM' is from~\cite{Blum:2014tka} (note that this 
uses a different method to fix $Z_q$ than is used here),
`BMW clover RI-MOM' is from~\cite{Durr:2010vn, Durr:2010aw},
`HPQCD HISQ JJ + $m_c/m_s$' (blue) is 
from~\cite{McNeile:2010ji, Davies:2009ih}. 
The light grey shaded band 
shows the evaluation of $96(+8,-4)$ GeV from 
the Particle Data Group~\cite{pdg}. The dark grey shaded band 
gives the weighted average (allowing for correlations) of 
the $n_f=2+1+1$ results as described in the text.  } 
\label{fig:mscomp}
\end{figure}

Lattice QCD is the method of choice for determining quark masses because 
it gives direct access to those parameters in the QCD Lagrangian
and allows them to be tuned very 
cleanly against hadron masses measured in experiment. 
As emphasised in Section~\ref{sec:intro}, the key complication in determining quark 
masses is in providing the matching factor from the quark mass in a particular 
lattice QCD regularisation scheme to the preferred $\overline{\text{MS}}$ continuum 
regularisation scheme. Here we compare two accurate methods for providing this matching factor 
directly for the $c$ quark mass: one is to take the continuum limit of time-moments in the 
current-current correlator method and the 
second is to use an intermediate momentum-subtraction scheme whose definition on the 
lattice translates directly to the continuum. Both methods then use continuum 
QCD perturbation theory for the final matching step. 
The two methods are very different in approach; one uses gauge-invariant 
meson correlators (2-point functions) in position space that are 
extrapolated to the continuum limit before matching to perturbation 
theory; the other uses gauge-noninvariant 2-point and 3-point 
functions in momentum space, obtaining a renormalisation factor 
at each value of the lattice spacing. 
Both methods have mechanisms for testing and estimating systematic 
uncertainties 
within them and so both are capable of yielding a complete error budget 
for the final result.   
A comparison of the two methods is important to make sure that 
the uncertainties are being 
fully controlled. 
The best comparison in this respect is a direct one between the two different 
methods for the same lattice QCD quark formalism on the same gauge field configurations. 
This is the comparison that we provide here, for the first time. 

Both the current-current correlator method and the intermediate 
momentum-subtraction scheme approaches have variants that allow for improved control of 
systematic errors. Our comparison uses the best variant to date of each method.  
We compare earlier results from the improved current-current 
correlator method (method c) from~\cite{Chakraborty:2014aca} 
to those obtained here using the RI-SMOM intermediate 
scheme~\cite{Sturm:2009kb}, which improves on earlier 
momentum-subtraction schemes in having 
smaller nonperturbative and perturbative 
uncertainties\footnote{Both the JJ and RI-SMOM methods go beyond quark 
mass determination and can be used 
more generally for current renormalisation~\cite{Sturm:2009kb, Donald:2012ga}, widening the importance 
of providing a comparison of the two approaches.}. 

The lattice QCD results that we give here are for the matching factor, $Z_m^{\text{SMOM}}$, 
determined 
nonperturbatively on the lattice between the HISQ quark mass and that in the symmetric 
MOM (RI-SMOM) scheme at multiple different scales between 2 and 5 GeV. 
We obtained results at different 
valence (Section~\ref{subsec:valmass}) 
and sea (Section~\ref{subsec:seamass}) quark masses and different 
spatial volumes (Section~\ref{subsec:volume}) 
to understand in detail what the sources of lattice systematic uncertainty are. 
We find that variations with sea mass and volume are barely discernible over a large 
range and valence 
mass effects are small (in contrast to those seen in the RI-MOM intermediate scheme). 
We combine results for the matching factor at 3 different values of the lattice spacing 
with the tuned $c$ and $s$ HISQ masses at that value of the lattice spacing obtained 
from the HPQCD current-current correlator calculation~\cite{Chakraborty:2014aca}. 
We are then able, by including an SMOM to $\overline{\text{MS}}$ matching factor (where 
we include the impact of having a non-zero $c$ mass in the sea), 
to extrapolate the resulting masses 
in the $\overline{\text{MS}}$ scheme to the $a=0$ continuum limit (see Section~\ref{sec:mass}).  
By having results at a range of values of the scale, $\mu$, we are able to 
include a systematic uncertainty for remaining nonperturbative (condensate) contributions 
that depend on inverse powers of $\mu$. These are expected to be much smaller than 
in the RI-MOM case, but we demonstrate their existence at several points 
in our calculation and they cannot be ignored. We find a residual effect 
of 0.2~\% in our error 
budget from these nonperturbative contributions 
(see Table~\ref{tab:errbudget}). 

The values we obtain at the reference scale of 3 GeV for the $c$ and $s$ quark masses are: 
\begin{eqnarray}
\label{eq:m3vals}
\overline{m}_c (3 \GeV, n_f=4) &=& 0.9896(61)\, \mathrm{GeV} \\
\overline{m}_s (3 \GeV, n_f=4) &=& 0.08536(85)\, \mathrm{GeV} \nonumber
\end{eqnarray}
These results are to be compared to those from the current-current correlator 
method~\cite{Chakraborty:2014aca} using 
the same formalism on the same gluon field configurations. The $\overline{m}_c$ value 
of 0.9851(63) GeV is obtained directly in the current-current 
correlator method; the value of $m_s$ of 0.0845(7) GeV 
(adjusting the value 
quoted in~\cite{Chakraborty:2014aca} from $n_f=3$ to 4) 
uses 
in addition a fully nonperturbative determination of the mass ratio 
$m_c/m_s$~\cite{Chakraborty:2014aca}. 
There is good agreement between the two sets of results, within the
uncertainties quoted\footnote{Note that the $Z_m$ values quoted in~\cite{Chakraborty:2014aca} 
cannot be directly compared to those given here because they are defined to incorporate 
lattice spacing artefacts in a different way.}. 
The uncertainties are small (~1\%) in both approaches
and they are not strongly correlated between them. This is because the dominant 
sources of uncertainty are very different in the two cases: for the RI-SMOM calculation 
a key source of error is that from the determination of the tuned lattice quark masses, 
whereas the current-current correlator method is less sensitive to those and 
a larger source of uncertainty is that 
from missing higher order terms in the continuum QCD perturbation theory for that quantity. 
Note that $\alpha_s$ can be determined in the same calculation in the current-current 
correlator approach and the correlation between $\alpha_s$ and $\overline{m}_c$ 
determined~\cite{Chakraborty:2014aca}. For this RI-SMOM calculation we must take 
a value of $\alpha_s$ from elsewhere when we need one for the SMOM to $\overline{\mathrm{MS}}$ 
conversion.  
Both the JJ and RI-SMOM mass determinations have uncertainties from lattice 
discretisation effects but they appear 
in different ways: in the RI-SMOM method it is the quark mass itself that is extrapolated 
to the continuum limit; in the current-current correlator method it is the time-moments of 
the correlator that are extrapolated. The agreement between the two methods is then 
a strong additional indication that 
the separate sources of systematic error are well controlled.   

We can run our new RI-SMOM values for $\overline{m}_c$ and $\overline{m}_s$ 
given in eq.~(\ref{eq:m3vals}) to other scales for comparison 
with other results. The scale used for the $\overline{m}_c$ is often that of 
$\overline{m}_c$ itself and for $\overline{m}_s$, 2 GeV~\cite{pdg}. 
We give these values below using 4-loop perturbative QCD running 
in the $\overline{\text{MS}}$ scheme to run down from the higher scales 
at which the masses were determined: 
\begin{eqnarray}
\label{eq:m-canon-vals}
\overline{m}_c (\overline{m}_c, n_f=4) &=& 1.2757(84) \, \mathrm{GeV}, \\
\overline{m}_s (2 \,\text{GeV}, n_f=4) &=& 0.09449(96) \, \mathrm{GeV}. \nonumber
\end{eqnarray}
The uncertainty on the value of both $\overline{m}_c$ and $\overline{m}_s$ 
increase because of the uncertainty 
in $\alpha_s$ at these low scales. Quoting $\overline{m}_c$ at its own scale 
reduces the resulting error a little because the mass runs down as its scale 
goes up. 
The comparable results from the current-current correlator method are 1.2715(95) GeV 
for $\overline{m}_c$ and 0.0936(8) GeV for $\overline{m}_s$~\cite{Chakraborty:2014aca}
{\footnote{Note 
that there is a typographical error in that work so that the value 
quoted at 2 GeV is for $n_f=4$ and not $n_f=3$ as stated.}}. 

The information that the JJ and RI-SMOM methods agree means that an average of the two results 
should have a reduced uncertainty. We must take care in the averaging to allow for 
the correlations between the two methods and we do this by dividing the 
uncertainty in each case into correlated and uncorrelated pieces and then fitting 
the two results to a constant. The correlated 
portion comes from the tuning of the quark mass and determination of the lattice 
spacing (as given in the error budget) and is taken to be 100\% correlated between 
the two methods.  The breakdown in the uncertainty is then:
\begin{eqnarray}  
\label{eq:errbdown}
\overline{m}_c(3\,\mathrm{GeV})&:& \mathrm{JJ}: \, 0.9851(22)(59) \,\mathrm{GeV}\\
&& \mathrm{SMOM}:\, 0.9896(32)(52) \, \mathrm{GeV}, \nonumber \\
\overline{m}_s(3\,\mathrm{GeV})&:& \mathrm{JJ}:\,  0.0845(4)(6) \, \mathrm{GeV} \nonumber \\
&& \mathrm{SMOM}:\, 0.08536(71)(47) \,\mathrm{GeV} \nonumber
\end{eqnarray}
with the first uncertainty being correlated and the second uncorrelated. 
This corresponds to a correlation coefficient between the two results for 
$m_c$ of 0.18 and for $m_s$ of 0.47. 
The resulting averages are given below. The $Q$ value for the fits 
were 0.6 for $\overline{m}_c$ and 0.3 for $\overline{m}_s$.
\begin{eqnarray}
\label{eq:av}
\langle\overline{m}_c(3\,\mathrm{GeV})\rangle_{\text{JJ,SMOM}} &=& 0.9874(48)\,\mathrm{GeV}, \\
\langle\overline{m}_s(3\,\mathrm{GeV})\rangle_{\text{JJ,SMOM}} &=& 0.08478(65)\,\mathrm{GeV}. \nonumber
\end{eqnarray}
The uncertainties are reduced by a small amount over those of the two separate values, 
giving a 0.8\% uncertainty in $\overline{m}_s$ and a 0.5\% uncertainty in 
$\overline{m}_c$

We can also run these averages down to the lower scales discussed above to give 
\begin{eqnarray}
\label{eq:avlow}
\langle\overline{m}_c(\overline{m}_c,n_f=4)\rangle_{\text{JJ,SMOM}} &=& 1.2737(77)\,\mathrm{GeV}, \\
\langle\overline{m}_s(2\,\mathrm{GeV}, n_f=4)\rangle_{\text{JJ,SMOM}} &=& 0.09385(75)\,\mathrm{GeV}. \nonumber
\end{eqnarray}

\subsection{World average for $\overline{m}_c(\overline{m}_c)$}
\label{subsec:worldmc}
In Figure~\ref{fig:mccomp} we compare our results for $\overline{m}_c(\overline{m}_c,n_f=4)$ 
to previous lattice QCD results. This graphically shows the agreement between 
the RI-SMOM results here and the directly comparable current-current correlator results 
(the top and third from top values in the figure). 
We also include a result (second from the top in the figure) 
obtained recently by the Fermilab Lattice, 
MILC and TUMQCD collaborations from a new method they have developed 
using the minimal renormalon-subtracted (MRS) scheme~\cite{Brambilla:2017hcq, Bazavov:2018omf}. This 
uses Heavy Quark Effective Theory to map out the heavy quark mass dependence of 
pseudoscalar heavy-light meson masses calculated in lattice QCD. 
A well-defined quark mass for this expansion is obtained by 
identifying and removing the leading renormalon from the perturbative expansion 
for the pole mass in terms of the $\overline{\text{MS}}$ mass. 
This MRS mass then has available a high-order continuum QCD perturbative matching 
to the $\overline{\mathrm{MS}}$ scheme. 
Application of this method also yields 1\%-accuracy and agrees, well within its 
uncertainty, with our JJ and RI-SMOM results. 

Although the majority of results in Figure~\ref{fig:mccomp} have been obtained 
using the HISQ formalism there are results using other formalisms that demonstrate 
good agreement, for example the results from JLQCD using domain-wall quarks~\cite{Nakayama:2016atf}. 
The results are divided into those obtained on gluon 
field configurations that include $u$, $d$, $s$ and $c$ quarks in the sea (as here) and 
those that include $u$, $d$ and $s$ quarks in the sea. Results obtained on gluon field 
configurations that include only $u$ and $d$ quarks in the sea are not shown, because it 
is not clear how to connect them in perturbative QCD (adding an $s$ sea quark) 
to the values shown here.  
We see good agreement between almost all the results. The majority of accurate previous results
used the current-current correlator method; the RI-MOM intermediate scheme has larger 
sources of systematic error for the reasons discussed in Section~\ref{sec:intro}.  
The current-current correlator results are tagged with `a', `b' or `c' to denote 
different implementations of the method; this will be discussed further below. 

We can provide a new world average for $\overline{m}_c(\overline{m}_c)$ allowing 
for correlations between the two HPQCD results by using the average given 
in eq.~(\ref{eq:avlow}) 
and then combining in a weighted average with the ETMC result~\cite{Carrasco:2014cwa}
and the Fermilab Lattice/MILC/TUMQCD result~\cite{Bazavov:2018omf}. 
The ETMC result (using the RI-MOM approach in the Twisted Mass 
formalism), 1.348(42) GeV, is nearly 2$\sigma$ from the HPQCD results and uncorrelated with it. 
The Fermilab Lattice/MILC/TUMQCD result is 1.273(10) GeV and is correlated with 
the HPQCD JJ result because it uses the HPQCD determination of $\alpha_s$ obtained 
concurrently with $m_c$ in~\cite{Chakraborty:2014aca}. The correlation coefficient between
$\alpha_s$ and $m_c$ is given there as 0.16. Since the uncertainty in the Fermilab 
Lattice/MILC/TUMQCD result is strongly dominated by that from $\alpha_s$ we apply 
a correlation coefficient of 0.16 between that result and the HPQCD average. 
The HPQCD average will be slightly less correlated with it than the JJ 
result alone. This allows, however, for some further correlation through the fact that 
all of these calculations use some of the same sets of gluon field configurations 
and are done with the same quark formalism, although different lattice QCD quantities 
are calculated and the lattice spacing was fixed and quark masses tuned in a different way. 
 
The weighted average of lattice QCD results including 4 flavours of sea quarks 
is then 
\begin{equation}
\label{eq:cav}
\overline{m}_c(\overline{m}_c,n_f=4)_{\mathrm{2+1+1 \,av.}} = 1.2753(65)\,\mathrm{GeV},
\end{equation}
shown as the dark shaded band on Figure~\ref{fig:mccomp}. 
The average has a poor $\chi^2/\mathrm{dof}$ of 3 because of the tension between 
the ETMC result, which has almost no impact on the average, and the other three values. 
Our dark shaded band is a somewhat 
narrower band than the evaluation given in the Particle Data Tables~\cite{pdg}, 
shown as the lighter shaded band. 

\subsection{World average for $\overline{m}_s(2\,\mathrm{GeV})$}
\label{subsec:worldms}
Figure~\ref{fig:mscomp} provides a comparison of lattice QCD results for 
the $s$ quark mass in the $\overline{\mathrm{MS}}$ scheme at a scale of 2 GeV. 
Results are given for calculations with 4 flavours in the sea (as here) and 
also 3 flavours in the sea.  There is very little difference 
(0.2 MeV, with $n_f$=3 larger) between 
these from perturbative QCD. Again results for 2 flavours in the sea are not shown 
since they cannot be connected perturbatively to the more realistic 3 and 4 flavour 
results. 
There is reasonably good overall agreement between results using current-current 
correlator methods or the MRS scheme and $m_c/m_s$ ratios and those using RI-MOM and RI-SMOM 
intermediate schemes direcetly for $m_s$, as well as between results using a variety of quark formalisms. 
Our new RI-SMOM result, however, shows a 2.7$\sigma$ tension with the earlier 
RI-SMOM result from RBC/UKQCD~\cite{Blum:2014tka} using domain-wall quarks. 
In~\cite{Blum:2014tka} the RI-SMOM implementation is slightly different, using 
the vector current vertex to fix $Z_q$. Two values of the lattice spacing are 
used but only one value of $\mu$ (3 GeV) and possible nonperturbative effects 
are not included in the analysis or allowed for as a systematic uncertainty. 

We can determine a new world average for the results from 4 flavour calculations 
allowing for correlations between the two different HPQCD results and the 
Fermilab/MILC/TUMQCD MRS result.  We combine them at 3 GeV where we have correlated 
the breakdown of errors in the HPQCD results in eq.~(\ref{eq:errbdown}). There 
we take the uncertainties associated with fitting/scale-setting from 
the Fermilab/MILC/TUMQCD result to be 100\% correlated with the HPQCD results 
(an overestimate given the different methods used) but take $\alpha_s$ and 
statistical uncertainties to be uncorrelated, ignoring the relatively small 
correlations between the uncertainty coming from $\alpha_s$ (which does not dominate 
in this case) and a part of the HPQCD JJ errors coming from $m_c$.  
We combined this with the uncorrelated ETMC result run to 3 GeV. 
This average gives 0.08393(43) GeV with a $\chi^2/\mathrm{dof}$ of 2.5. 
Inflating the uncertainty by $\sqrt{2.5}$ to take account of this 
more general tension, and running down to 2 GeV 
gives 
\begin{equation}
\label{eq:sav}
\overline{m}_s(2\,\mathrm{GeV},n_f=4)_{\mathrm{2+1+1 \,av.}} = 0.09291(78)\,\mathrm{GeV} .
\end{equation}
This is given as the dark shaded band in Figure~\ref{fig:mscomp} to be compared 
to the light shaded band of the evaluation in the Particle Data Tables~\cite{pdg}. 
The Particle Data Tables result seems unduly pessimistic about our level of knowledge of 
the $s$ quark mass and has a high central value, given the accuracy of lattice QCD results
now available. 

\subsection{Future}
\label{subsec:future}
Accurate though these results are, it is worth asking what the prospects are 
for reducing uncertainties further in the future. 
The original current-current correlator method (method a)~\cite{Allison:2008xk} 
used lattice QCD results for quarks tuned to the $c$ mass only and 
so one of the largest sources of uncertainty was from missing higher order terms 
in the perturbative series for time-moments because the scale of $\alpha_s$ was 
related to $m_c$. Subsequently (in method b)~\cite{McNeile:2010ji} heavier quark masses 
were included, giving access to $m_b$ but also reducing the perturbative 
uncertainty because the combined fit now included $\alpha_s$ terms evaluated at higher scales. 
The newest variant, method c~\cite{Chakraborty:2014aca} also includes results for 
quarks with heavier masses than $c$. By making use of quark mass ratios, 
$m_c$ is then determined through a perturbative series for the time-moments 
in which the scale of $\alpha_s$ is set by the heavier quark masses. 
This method then offers the potential to reduce the perturbative uncertainty by 
working on yet finer lattices where a given value of quark mass in lattice units 
corresponds to a higher quark mass. This will also reduce the sizeable uncertainty from 
the $a \rightarrow 0$ extrapolation. 

For the RI-SMOM intermediate scheme method used here, the largest sources of 
uncertainty are those from the bare tuned lattice quark masses which in turn 
depend on the determination of the lattice spacing. Working on finer lattices could
cut this uncertainty significantly since the lattice spacing is fixed 
from $f_{\pi}$ in the continuum and physical $u/d$ mass limit~\cite{fkpi}. The impact of 
missing higher order 
terms in the $\mathrm{SMOM}$ to $\overline{\mathrm{MS}}$ matching would be reduced 
by going to higher $\mu$ values and this is also possible on finer lattices. 
At the same time this would also reduce the impact of nonperturbative contributions since they 
fall rapidly with $\mu$. 

In conclusion, lattice QCD now has three very different
methods for 
determination of quark masses at an accuracy of 1\% or better. 
They yield consistent results for $\overline{m}_c(\overline{m}_c)$ and 
$\overline{m}_s(2 \mathrm{GeV})$. We have given here 
a particularly strong test of the consistency of the current-current correlator 
and RI-SMOM methods.  
Lattice QCD calculations are then well on the way to providing the 
accuracy needed for stringent future tests of the Standard Model. 

\subsection*{\bf{Acknowledgements}} 

We are grateful to the MILC collaboration for the use of 
their configurations and their code. 
We thank E. Follana and E. Royo-Amondarain for gauge-fixing the 
superfine configurations. 
Computing was done on the Darwin supercomputer at the University of 
Cambridge High Performance Computing Service as part of the DiRAC facility, 
jointly funded by the Science and Technology Facilities Council, 
the Large Facilities Capital Fund of BIS and 
the Universities of Cambridge and Glasgow. 
We are grateful to the Darwin support staff for assistance. 
Funding for this work came from the 
National Science Foundation, the Royal Society, 
the Science and Technology Facilities Council 
and the Wolfson Foundation.

\appendix 

\section{Perturbative matching from SMOM to {\boldmath{$\MSbar$}} for 
non-zero sea quark mass\label{appendix:seamass}}
\begin{figure}
\includegraphics[width=0.45\textwidth]{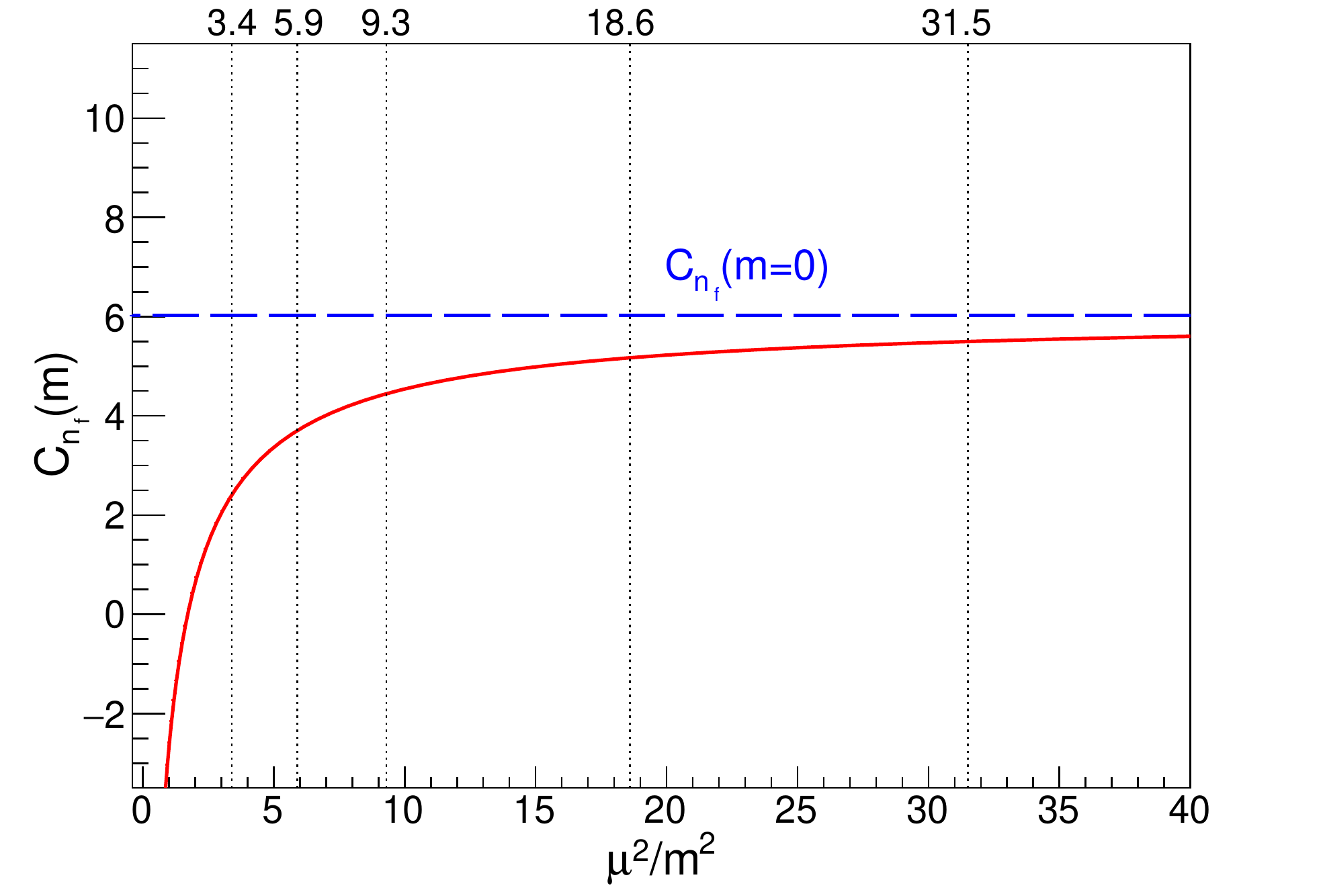}
\caption{ The coefficient $C_{n_f}$ which enters the
  $Z_m^{\MSbarSMOM}$ factor at
  $\mathcal{O}(\alpha_s^2)$ as a function of $\mu^2/m^2$ for a
  massive sea quark of mass $m$. The dashed, blue line shows the $m=0$
  result. The solid, red line shows the exact result of a numerical evaluation of
  the contribution which is valid across the full range of $m$ values. The
  vertical lines denote the location of the different $\mu^2/m^2$
  values, which are used in this section.\label{fig:massmatch}}
\end{figure}
We have defined our SMOM scheme to have massive sea quarks so that we can work 
at physical sea quark masses. We then need a perturbative matching from 
SMOM to the $\MSbar$ scheme that allows for massive sea quarks. 
Sea quarks do not appear in the matching calculation until $\mathcal{O}(\alpha_s^2)$ 
and even then they contribute only a very small part to that coefficient. 
Given the very small changes that we expect, we need only to consider the 
case of the most massive sea quark that we have, i.e. the $c$ quark. We work close to the 
physical mass for $c$ and this in turn (in the $\MSbar$ scheme) 
is a sizeable fraction of $\mu$ for the $\mu$ values that we use. 
We might therefore expect the $\alpha_s^2$ coefficient from 
the $c$ quark in the sea to be significantly different from that for a massless 
$c$ quark.  
Our results show that indeed this is true for $\mu$ close to our lower value 
of 2~GeV 
but even so this has very little impact on $Z_m^{\MSbarSMOM}$.

The two-loop coefficient $c_2$ of eq.~(\ref{eq:Zmpert}) can be
decomposed into two terms: a contribution which is free of internal
sea quarks, $C_{n_f=0}$, and a contribution which depends on
them, $C_{n_f}(m)$. It can be written as
\begin{equation}
\label{eq:Cmdef}
c_2=\frac{1}{(4\pi)^2}\left[C_{n_f=0}+C_FT_Fn_f C_{n_f}(m)\right],
\end{equation}
where $C_F=4/3$ and $T_F=1/2$ are the usual colour factors.  The
two-loop coefficient $C_{n_f=0}$ of
$Z_m^{\MSbarSMOM}$ as well as the piece
$C_{n_f}(m=0)$ which is proportional to the number of massless sea
quarks has been determined in~\cite{Gorbahn:2010bf,Almeida:2010ns}.  The
result for $C_{n_f}(m=0)$ reads
\begin{equation}
\label{eq:Cnfm0}
C_{n_f}(m=0) = \left[ \frac{83}{6} + \frac{40}{27}\pi^2 - \frac{20}{9}\Psi^{\prime}\!\left(\tfrac{1}{3}\right)\right]=6.020...,
\end{equation}
where $\Psi^{\prime}$ is the derivative of the digamma function.  For a
massive internal quark the result for $C_{n_f}(m\neq 0)$ depends on the
ratio $\mu^2/m^2$.  For very heavy quarks $m\gg\mu$ the result for
$C_{n_f}(m\neq 0)$ can be obtained in terms of an expansion in
$\mu^2/m^2$ where the leading term reads
\begin{eqnarray}
C_{n_f}(m)&=&
- \frac{89}{18}
+ \log\!\left(\tfrac{\mu^2}{m^2}\right)\*\!
\left[
    \frac{26}{3} 
  + \frac{8}{9}\*\pi^2 
  - \frac{4}{3}\*\Psi^{\prime}\!\left(\tfrac{1}{3}\right)
\right]
\nonumber\\
&-& 2 \*\log^2\!\left(\tfrac{\mu^2}{m^2}\right)
           +\mathcal{O}\!\left(\tfrac{\mu^2}{m^2}\right)\;\mbox{for}\;m\to\infty.
\label{eq:Cminfty}
\end{eqnarray}
It can be derived with the help of the QCD decoupling
functions. In order to access the complete mass dependence, 
we also have calculated the exact result, which
is valid in any mass region. It is plotted in
Figure~\ref{fig:massmatch}.  We have checked by computing power
corrections in ${\mu^2/m^2}$ to eq.~(\ref{eq:Cminfty}) that the
expansion coincides as expected with the exact result for large values
of $m$.

If we take $\overline{m}_c(3~\mathrm{GeV})$
from~\cite{Chakraborty:2014aca}, then the value of the ratio
$\mu^2/{\overline{m}}^2(\mu)$ for $\mu$ = 2, 2.5, 3, 4 and 5~GeV reads
$\mu^2/{\overline{m}}^2(\mu)= 3.4$, 5.9, 9.3, 18.6 and 31.5. Table~\ref{tab:deltaCm}
gives the resulting shifts $\Delta C_m$ in $C_{n_f}$ from the $m=0$
result to the result that we need, which includes a massive $c$ quark, 
at these five values of $\mu$.
 
\begin{table}
\caption{Changes to the $\alpha_s^2$ coefficient of the perturbative
matching factor $Z_m^{\MSbarSMOM}$ as a
result of having a massive $c$ quark in the sea.  The second
column gives the change in $C_{n_f}$, i.e. $\Delta
C_m=C_{n_f}(m)-C_{n_f}(m=0)$ (for $n_f=1$); the third column gives the change
in the $\alpha_s^2$ coefficient $c_2$ (eq.~(\ref{eq:Zmpert}) and
Table~\ref{tab:mom-msbar}) for five different values of the
scale parameter, $\mu$, which is given in the first column. In the fourth
column we give values for $\alpha_s$ in the
$\MSbar$ scheme at each value of $\mu$. These are
obtained by four-loop running in the $\MSbar$ scheme
from a value of 0.2128(25) at a scale of 5~GeV with
$n_f=4$~\cite{Chakraborty:2014aca}.\label{tab:deltaCm}}
\begin{ruledtabular}
\begin{tabular}{llll}
$\mu$ (GeV) & $\Delta C_m$ & $\Delta c_2$ & $\alpha_{\MSbar}(\mu,n_f=4)$ \\
\hline 
2 & -3.6 & -0.015 & 0.3030(54) \\ 
2.5 & -2.3 & -0.010 & 0.2741(43) \\ 
3 & -1.6 & -0.007 & 0.2545(37)  \\
4 & -0.9 & -0.004 & 0.2291(29)  \\
5 & -0.5 & -0.002 & 0.2128(25)  \\
\end{tabular}
\end{ruledtabular}
\end{table}
 
We also give the resulting shift $\Delta c_2$ in $c_2$ of
eq.~(\ref{eq:Cmdef}).  We can see from Table~\ref{tab:deltaCm} that the
only significant effect is for $\mu$ = 2~GeV where the shift is about
40\% of the $\alpha_s^2$ coefficient.  This coefficient is very small,
however, and so the impact of this shift is also very small, less than
about -0.2\% on $Z_m^{\MSbarSMOM}$.  We use
the results in Table~\ref{tab:deltaCm} to shift the values of
$Z_m^{\MSbarSMOM}$ used in order to convert our SMOM
results to the $\MSbar$ scheme in
Section~\ref{sec:mass}.

\section{OPE for the scalar vertex operator} 
\label{appendix:ope}

The operator 
\begin{eqnarray}
\label{Sopdef}
\hat{\Lambda}_S &&\equiv -\int d^4x d^4 y e^{ip^{\prime}\cdot y - i p\cdot x} \\
&& T\left[ \overline{\psi}_b(0)\psi_b(0)\frac{1}{12}\Tr\left(S^{-1}(p^{\prime})\psi_a(y)\overline{\psi}_a(x)S^{-1}(p)\right)\right] \nonumber
\end{eqnarray}
has vacuum expectation value $\Tr \Lambda_S(p,p^{\prime})/12$ that we use to define $Z_S$ 
in eq.~(\ref{eq:ZSdef}). Here $a$ and $b$ are colour indices (summed over). 
For the symmetric kinematic configuration that we use, $p^2=(p^{\prime})^2=(p-p^{\prime})^2=-\mu^2$ 
with $\mu$ large, the operator is truly short-distance and has an 
OPE expansion in terms of local operators 
of increasing dimension
multiplying inverse powers of $\mu$: 
\begin{equation}
\label{eq:opedef}
\hat{\Lambda}_S = c_1(\mu) \mathbf{1} + c^{\Lambda_S}_{A^2}(\mu)\frac{A^2}{\mu^2} + \ldots 
\end{equation}
On taking the vacuum expectation, the first term yields the perturbative expansion 
and the second and higher terms give power-suppressed nonperturbative contributions. 
To determine the coefficients of these latter terms, matrix elements of $\hat{\Lambda}_S$
can be taken between states for which the $\mathbf{1}$ operator gives zero. 
For example, the scattering amplitude between low-momentum gluon fields with $k^2 \rightarrow 0$ 
and $k\cdot\varepsilon = 0$ can be evaluated for both sides of the OPE. 
The left-hand side gives 
\begin{eqnarray}
\label{eq:lambda1loop}
\langle \hat{\Lambda}_S \rangle &=& -\pi\alpha_s C_F \int_k \Tr\left[ \gamma_{\rho}\frac{i}{\Xsl{p^{\prime}}-\Xsl{k}-m}\frac{i}{\Xsl{p}-\Xsl{k}-m}\gamma_{\sigma}\right] \varepsilon_1^{\rho}\varepsilon_2^{\sigma} \nonumber \\
&=& \frac{\pi\alpha_s C_F}{\mu^4} \Tr\left[\Xsl{\varepsilon_1}\Xsl{p^{\prime}}\Xsl{p}\Xsl{\varepsilon_2}\right] \nonumber \\
&=& -\frac{2\pi\alpha_sC_F}{\mu^2} \varepsilon_1 \cdot \varepsilon_2 
\end{eqnarray}
on averaging over directions of 
the external momenta. 
The right-hand side gives 
\begin{equation}
\label{eq:opeoneloop}
\langle OPE \rangle = \frac{8 c^{\Lambda_S}_{A^2}}{\mu^2} \varepsilon_1 \cdot \varepsilon_2 
\end{equation}
Hence 
\begin{equation}
\label{eq:cA2lambda}
c^{\Lambda_S}_{A^2} = - \frac{\pi \alpha_s C_F}{4}
\end{equation}
Note that it is clear from this that the pseudoscalar vertex would have the 
same coefficient for the leading condensate contribution and hence this will 
vanish from the difference $\Lambda_S - \Lambda_P$, as we illustrate 
in Figure~\ref{fig:ZPS}.  

A parallel analysis can be done for the operator 
\begin{eqnarray}
\label{eq:sigmadef}
\hat{\Sigma} \equiv &&\frac{-i}{p^2}\int d^4 x e^{ip\cdot x} \\
&& T\left[ \frac{1}{12}\Tr\left( \Xsl{p}S^{-1}(p)\psi_a(x)\overline{\psi}_a(0)S^{-1}(p) \right)\right] \nonumber 
\end{eqnarray}
whose vacuum expectation value is $\Tr(\Xsl{p}S^{-1}(p))/(12p^2)$. We use this to define $Z_q$ 
in eq.~(\ref{eq:Zq}).  
Scattering from low-momentum gluons gives 
\begin{eqnarray}
\langle \hat{\Sigma}\rangle &=& - \frac{i\pi\alpha_s C_F}{p^2}\int_k \Tr\left[ \Xsl{p}\gamma_{\rho}\frac{i}{\Xsl{p}-\Xsl{k}-m}\gamma_{\sigma}\right] \varepsilon_1^{\rho} \varepsilon_2^{\sigma} \nonumber \\
&=& \frac{2\pi\alpha_s C_F}{\mu^2} \varepsilon_1 \cdot \varepsilon_2
\end{eqnarray}
Equating this to the result from the matrix element of the OPE, 
$(8c^{\Sigma}_{A^2}/\mu^2)\varepsilon_1 \cdot \varepsilon_2 $, 
gives 
\begin{equation}
\label{eq:cA2sigma}
c^{\Sigma}_{A^2} =  \frac{\pi \alpha_s C_F}{4}
\end{equation}
in agreement with results (for the expansion of the quark propagator) in~\cite{Lavelle:1988eg, Lavelle:1992yh, Chetyrkin:2009kh}. 

We use the ratio of vacuum matrix elements of $\hat{\Lambda}_S$ and $\hat{\Sigma}$ to 
define $Z_m$ (eq.~(\ref{eq:ZSdef})). Hence the leading condensate contribution in 
$Z_m$ has coefficient at $\mathcal{O}(\alpha_s)$ of $c^{\Lambda_S}_{A^2} - c^{\Sigma}_{A^2}$, 
i.e. $2\pi\alpha_s/3$.

\bibliography{paper}
\end{document}